\documentclass[12pt,a4paper]{article}
\usepackage{appendix}
\usepackage{ifthen}
\usepackage[lofdepth,lotdepth]{subfig}
\usepackage{amsmath}
\usepackage{xspace}
\usepackage{epsfig}
\usepackage{mciteplus}
\usepackage{rotating}
\usepackage{dcolumn}
\usepackage{url}
\usepackage{hyperref}
\usepackage[numbers,sort&compress]{natbib}
\usepackage{hypernat}
\hypersetup{ pdfborder=0 0 0, urlbordercolor=1 0 0 }

\usepackage{graphicx}
\usepackage{color}
\usepackage{colortbl}
\newboolean{uprightparticles}
\setboolean{uprightparticles}{false} 
\newboolean{articletitles}
\setboolean{articletitles}{true} 

\def\ux85 {UX85\xspace}

\def\velo   {VELO\xspace}

\ifthenelse{\boolean{uprightparticles}}%
{

 \def\PDelta      {\ensuremath{\Delta}\xspace}                 
 \def\PXi      {\ensuremath{\Xi}\xspace}                 
 \def\PLambda      {\ensuremath{\Lambda}\xspace}                 
 \def\PSigma      {\ensuremath{\Sigma}\xspace}                 
 \def\POmega      {\ensuremath{\Omega}\xspace}                 
 \def\PUpsilon      {\ensuremath{\Upsilon}\xspace}                 
 
 \def\PB      {\ensuremath{\mathrm{B}}\xspace}                 
                  
 \def\PD      {\ensuremath{\mathrm{D}}\xspace}

 \def\PK      {\ensuremath{\mathrm{K}}\xspace}

 \def\Pi      {\ensuremath{\mathrm{i}}\xspace}

}
{

 \mathchardef\PDelta="7101
 \mathchardef\PXi="7104
 \mathchardef\PLambda="7103
 \mathchardef\PSigma="7106
 \mathchardef\POmega="710A
 \mathchardef\PUpsilon="7107
                  
 \def\PB      {\ensuremath{B}\xspace}                 
                  
 \def\PD      {\ensuremath{D}\xspace}

 \def\PK      {\ensuremath{K}\xspace}

 \def\Pi      {\ensuremath{i}\xspace}

}

\def\kaon  {\ensuremath{\PK}\xspace}
  \def\Kbar  {\kern 0.2em\overline{\kern -0.2em \PK}{}\xspace}

\def\Kz    {\ensuremath{\kaon^0}\xspace}
\def\Kzb   {\ensuremath{\Kbar^0}\xspace}
\def\KzKzb {\ensuremath{\Kz \kern -0.16em \Kzb}\xspace}
\def\Kp    {\ensuremath{\kaon^+}\xspace}
\def\Km    {\ensuremath{\kaon^-}\xspace}

\def\KpKm  {\ensuremath{\Kp \kern -0.16em \Km}\xspace}
\def\KS    {\ensuremath{\kaon^0_{\rm\scriptscriptstyle S}}\xspace}

  \def\Dbar    {\kern 0.2em\overline{\kern -0.2em \PD}{}\xspace}
\def\D       {\ensuremath{\PD}\xspace}

\def\Dz      {\ensuremath{\D^0}\xspace}
\def\Dzb     {\ensuremath{\Dbar^0}\xspace}
\def\DzDzb   {\ensuremath{\Dz {\kern -0.16em \Dzb}}\xspace}
\def\Dp      {\ensuremath{\D^+}\xspace}
\def\Dm      {\ensuremath{\D^-}\xspace}

\def\DpDm    {\ensuremath{\Dp {\kern -0.16em \Dm}}\xspace}

  \def\Bbar    {\kern 0.18em\overline{\kern -0.18em \PB}{}\xspace}

  \def\Y#1S{\ensuremath{\PUpsilon{(#1S)}}\xspace}

\def\L {\ensuremath{\PLambda}\xspace}

\def\AT#1     {\ensuremath{A_{\mathrm{T}}^{#1}}\xspace}           

\def\C#1      {\ensuremath{\mathcal{C}_{#1}}\xspace}                       
\def\Cp#1     {\ensuremath{\mathcal{C}_{#1}^{'}}\xspace}                    
\def\Ceff#1   {\ensuremath{\mathcal{C}_{#1}^{\mathrm{(eff)}}}\xspace}        
\def\Cpeff#1  {\ensuremath{\mathcal{C}_{#1}^{'\mathrm{(eff)}}}\xspace}       
\def\Ope#1    {\ensuremath{\mathcal{O}_{#1}}\xspace}                       
\def\Opep#1   {\ensuremath{\mathcal{O}_{#1}^{'}}\xspace}                    

\newcommand{\tev}{\ensuremath{\mathrm{\,Te\kern -0.1em V}}\xspace}
\newcommand{\gev}{\ensuremath{\mathrm{\,Ge\kern -0.1em V}}\xspace}
\newcommand{\mev}{\ensuremath{\mathrm{\,Me\kern -0.1em V}}\xspace}
\newcommand{\kev}{\ensuremath{\mathrm{\,ke\kern -0.1em V}}\xspace}
\newcommand{\ev}{\ensuremath{\mathrm{\,e\kern -0.1em V}}\xspace}
\newcommand{\gevc}{\ensuremath{{\mathrm{\,Ge\kern -0.1em V\!/}c}}\xspace}
\newcommand{\mevc}{\ensuremath{{\mathrm{\,Me\kern -0.1em V\!/}c}}\xspace}
\newcommand{\gevcc}{\ensuremath{{\mathrm{\,Ge\kern -0.1em V\!/}c^2}}\xspace}
\newcommand{\gevgevcccc}{\ensuremath{{\mathrm{\,Ge\kern -0.1em V^2\!/}c^4}}\xspace}
\newcommand{\mevcc}{\ensuremath{{\mathrm{\,Me\kern -0.1em V\!/}c^2}}\xspace}

\def\mm   {\ensuremath{\rm \,mm}\xspace}

\def\gsim{{~\raise.15em\hbox{$>$}\kern-.85em
          \lower.35em\hbox{$\sim$}~}\xspace}
\def\lsim{{~\raise.15em\hbox{$<$}\kern-.85em
          \lower.35em\hbox{$\sim$}~}\xspace}

\newcommand{\mean}[1]{\ensuremath{\left\langle #1 \right\rangle}} 

\def\sqs   {\ensuremath{\protect\sqrt{s}}\xspace}

\def\pt         {\mbox{$p_{\rm T}$}\xspace}

\def\pythia     {\mbox{\textsc{Pythia}}\xspace}

\def\geant      {\mbox{\textsc{Geant4}}\xspace}

\def\tell1  {TELL1\xspace}
\def\ukl1   {UKL1\xspace}

\newcommand{\ie}{\mbox{\itshape i.e.}}

\def\phojet     {\mbox{\textsc{Phojet}}\xspace}
\def\photos      {\mbox{\textsc{Photos}}\xspace}

\begin{document}
\newcommand{\pp}{\ensuremath{\mbox{p-p}}}
\newcommand{\beq}[1]{\begin{equation}{\label{#1}}}
\newcommand{\eeq}[0]{\end{equation}}
\newcommand{\vev}[1]{{\langle #1 \rangle}}
\newcommand{\ptrans}{\ensuremath{p_{\rm T}}}
\newcommand{\Kshort}{\ensuremath{K^0_{\rm S}}}
\newcommand{\eq}[1]{eq.\ref{#1}}


\begin{titlepage}
\belowpdfbookmark{Title page}{title}

\pagenumbering{roman}
\vspace*{-1.5cm}
\centerline{\large EUROPEAN ORGANIZATION FOR NUCLEAR RESEARCH (CERN)}
\vspace*{1.5cm}
\hspace*{-5mm}\begin{tabular*}{16cm}{lc@{\extracolsep{\fill}}r}
\vspace*{-12mm}\mbox{\!\!\!\epsfig{figure=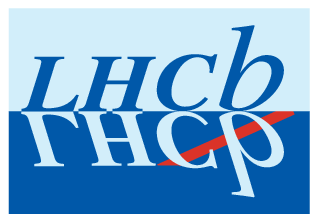,width=.12\textwidth}}& & \\
&& LHCb-PAPER-2011-011 \\
&& CERN-PH-EP-2011-209 \\
&& 14 December 2011 \\
\end{tabular*}
\vspace*{2cm}
\begin{center}
\Large
{\bf \boldmath
\huge {\boldmath Measurement of charged particle multiplicities
in $pp$ collisions at ${\sqs =7\tev}$ in the forward region}\\
\vspace*{1.5cm}
 }
\normalsize {
The LHCb Collaboration%
\footnote{Authors are listed on the following pages.}
}
\end{center}
\vspace{\fill}
\centerline{\bf Abstract}
\vspace*{5mm}\noindent
The charged particle production in proton-proton collisions is studied with the LHCb detector at a
centre-of-mass energy of ${\sqs = 7\tev}$ in different intervals of
pseudorapidity~$\eta$. The charged particles are reconstructed
close to the interaction region in the
vertex detector, which provides high
reconstruction efficiency in the $\eta$ ranges $-2.5<\eta<-2.0$ and
$2.0<\eta<4.5$. The data were taken with a minimum bias trigger, only
requiring one or more reconstructed tracks in the vertex detector.
By selecting an event sample with at least one track with a transverse momentum
greater than 1\gevc a hard QCD subsample is investigated.
Several event generators are compared with the data;
none are able to describe fully  the multiplicity distributions or the
charged particle density distribution as a function of $\eta$.
In general, the models underestimate the charged particle production.

\vspace*{2.cm}\noindent
{\em Keywords:} minimum bias, underlying event, particle
multiplicities, LHC, LHCb
\vspace{\fill}

\end{titlepage}
\newpage
\setcounter{page}{2}
\mbox{~}
\newpage

\belowpdfbookmark{LHCb author list}{authors}
\begin{flushleft}
\small
{\Large LHCb Collaboration}\\[4ex]
R.~Aaij$^{23}$,
C.~Abellan~Beteta$^{35,n}$,
B.~Adeva$^{36}$,
M.~Adinolfi$^{42}$,
C.~Adrover$^{6}$,
A.~Affolder$^{48}$,
Z.~Ajaltouni$^{5}$,
J.~Albrecht$^{37}$,
F.~Alessio$^{37}$,
M.~Alexander$^{47}$
% ...

{\footnotesize \it
% ...
$ ^{5}$Clermont Universit\'{e}, Universit\'{e} Blaise Pascal, CNRS/IN2P3, LPC, Clermont-Ferrand, France\\
$ ^{6}$CPPM, Aix-Marseille Universit\'{e}, CNRS/IN2P3, Marseille, France\\
% ...
$ ^{23}$Nikhef National Institute for Subatomic Physics, Amsterdam, The Netherlands\\
% ...
$ ^{35}$Universitat de Barcelona, Barcelona, Spain\\
$ ^{36}$Universidad de Santiago de Compostela, Santiago de Compostela, Spain\\
$ ^{37}$European Organization for Nuclear Research (CERN), Geneva, Switzerland\\
% ...
$ ^{42}$H.H. Wills Physics Laboratory, University of Bristol, Bristol, United Kingdom\\
% ...
$ ^{47}$School of Physics and Astronomy, University of Glasgow, Glasgow, United Kingdom\\
$ ^{48}$Oliver Lodge Laboratory, University of Liverpool, Liverpool, United Kingdom\\
% ...
$ ^{n}$LIFAELS, La Salle, Universitat Ramon Llull, Barcelona, Spain\\
% ...
}
\end{flushleft}

\mbox{~}\vfill
\noindent
{}
\vfill\mbox{~}

\cleardoublepage
\setcounter{page}{1}
\pagenumbering{arabic}

\section{Introduction}
The charged particle multiplicity is a basic observable that
characterizes the hadronic final state. The multiplicity distribution
is sensitive to the underlying QCD dynamics of the proton-proton
collision. ALICE~\cite{Aamodt:2010pp}, ATLAS~\cite{ATLAS:2010ir} and
CMS~\cite{Khachatryan:2010nk} have measured the charged multiplicity distributions
mainly covering the central region, while LHCb's geometrical acceptance allows
the dynamics of the collision to be probed in the forward region.
The forward region is in particular sensitive to low Bjorken-$x$ QCD dynamics and multi-parton
interactions (MPI)~\cite{Corke:2009tk}.

In this analysis, the charged particles are reconstructed in the vertex detector (\velo)
surrounding the interaction region. The \velo was designed to provide a uniform acceptance
in the forward region with additional coverage of the backward region. In the absence of almost
any magnetic field in the \velo region, the particle
trajectories are straight lines and therefore no acceptance
corrections as a function of momentum are needed. Since the \velo is close to
the interaction region, the amount of material before the particle
detection is small, minimising the corrections for particle
interactions with detector material.

This paper is organized as follows.
Section~\ref{sec:detector} gives a brief description of the LHCb
detector and the configuration used to record data in Spring 2010.
The Monte Carlo simulation and data selection are outlined
in Sections~\ref{sec:MonteCarlo} and~\ref{sec:selection} respectively,
with Section~\ref{sec:strategy} giving an overview of the analysis.
The systematic uncertainties are outlined in Section~\ref{sec:syserrors}.
The final results are discussed
in Section~\ref{sec:results} and compared with different model
expectations, before concluding in Section~\ref{sec:summary}.

\section{LHCb detector}
\label{sec:detector}

The LHCb detector is a single-arm magnetic dipole spectrometer with a
polar angular coverage with respect to the beam line of approximately
15 to 300~mrad in the horizontal bending plane, and 15 to 250~mrad in
the vertical non-bending plane.
The detector is described in detail
elsewhere~\cite{Alves:2008zz}.
A right-handed coordinate system is defined with its origin at the
nominal proton-proton interaction point, the $z$ axis along the beam
line and pointing towards the magnet, and the $y$ axis pointing upwards.

For the low luminosity running period of the LHC relevant for
this analysis, the probability of observing more than one
collision in a proton-proton bunch crossing ({\it pile-up}) is measured
to be $(3.7\pm0.4)\%$, dominated by a double interaction.
For the measurements presented in this paper the tracking detectors
are of particular importance.
The LHCb tracking system consists of the \velo
surrounding the proton-proton interaction region, a
tracking station (TT) before the dipole magnet, and three
tracking stations (T1--T3) after the magnet.  Particles
traversing from the interaction region to the downstream tracking
stations experience an integrated bending-field of approximately 4~Tm.

The \velo consists of silicon microstrip modules,
providing a measure of the radial and azimuthal coordinates,
$r$ and $\phi$, distributed in 23 stations arranged
along the beam direction.
The first two stations at the most upstream $z$ positions
are instrumented to provide
information on the number of visible interactions in the detector at the
first level of the trigger.
The \velo is constructed in two halves,
movable in the $x$ and $y$ directions so that it can be centered on the beam.
During stable beam conditions the two halves
are located at their nominal closed position, with active silicon at only
8~mm from the beams, providing full azimuthal coverage.

The TT station also uses silicon microstrip technology. The T1--T3
tracking stations have silicon microstrips in the region close
to the beam pipe, whereas straw tubes are employed in the outer region.

Though the particle multiplicity is measured using only tracks
reconstructed with the \velo,
momentum information is only available for ``long" tracks.
Long tracks are formed from hits in the \velo (before the magnet) and
in the T1--T3 stations (after the magnet).
If available, measurements in the TT station are added to the long track.

The LHCb trigger system consists of two levels. The first level is implemented in hardware and
is designed to reduce the event rate to a maximum of 1 MHz. The complete
detector is then read out and the data is sent to the second level, a software trigger.
For the early data taking period with
low luminosity used in this analysis a simplified trigger was used.
The first level  trigger was operated in pass-through mode.
A fast track reconstruction was performed in the software trigger
and events with at least one track observed in the VELO were accepted.

\section{Monte~Carlo simulation}
\label{sec:MonteCarlo}
Monte Carlo event simulation is used to correct for acceptance, resolution
effects and for background characterisation.
The detector simulation is based on the \geant~\cite{Agostinelli:2002hh} package. Details of the detector
simulation are given in Ref.~\cite{Alves:2008zz}.
The simulated material of the components of the VELO was compared with the
masses measured at the time of production and agreement was found to be within 15\%.
The Monte Carlo event samples are passed
through reconstruction and selection procedures identical to those for the data.

Elastic and inelastic proton-proton collisions are generated using
the \pythia 6.4 event generator~\cite{Sjostrand:2006za},
with CTEQ6L parton density functions~\cite{Pumplin:2002vw}, which
is tuned to lower energy hadron collider data~\cite{LHCb-PROC-2010-056}.
The inelastic processes include both single and double diffractive components.
The decay of the generated particles is carried out by EvtGen~\cite{Lange:2001uf},
with final state radiation handled by \photos~\cite{Golonka:2005pn}.
Secondary particles produced in material interactions are decayed through the \geant program.

\section{Data selection}
\label{sec:selection}
A sample of $3 \times 10^{6}$ events, collected during
May 2010, was used in this analysis. In order to minimize the contribution of
secondary particles and misreconstructed (fake) tracks, only the tracks
satisfying a set of minimal quality criteria are accepted.
To minimise fake tracks a cut on the $\chi^2$ per degree of freedom of the
reconstructed track, $\chi^{2}/{\rm ndf} < 5,$ is applied.
To further reduce fake tracks, and reduce duplicate tracks due to a split
of the reconstructed
trajectory, a cut of less than four missing \velo hits compared to the expectation is applied.
To ensure that tracks originate from the primary interaction,
the requirements $d_0 < 2 \, \mm$ and $z_0 < 3 \sigma_L$
are applied, where
$d_0$ is the track's closest distance to
the beam line, $z_0$ is the distance along the $z$ direction from the centre of the
luminous region and
$\sigma_L$ is the width of the luminous region extracted from
a Gaussian fit.

Tracks are considered for this analysis only if their pseudorapidity
is in either of the ranges $-2.5<\eta< -2.0$ and $2.0<\eta< 4.5$.
Pseudorapidity is defined as $-\ln[\tan(\theta/2)]$ where
$\theta$ is the polar angle of the particle with respect to the $z$ direction.
The forward range is divided in five equal sub-intervals with $\Delta\eta = 0.5$.

\section{Analysis strategy}
\label{sec:strategy}
The reconstructed multiplicity distributions
are corrected on an event by event basis to account for the tracking
and selection efficiencies and for the background contributions. These
corrected distributions are then used to measure the charged particle
multiplicities in each of the $\eta$ intervals (bins) through an unfolding
procedure.
Only events with tracks in the $\eta$ bins are included in the
distributions and subsequent normalisation.
The distributions are corrected for pile-up effects so they
represent the charged particle multiplicities, ${\rm n_{ch}}$, for single
proton-proton interactions.
No unfolding procedure is required for the charged particles pseudorapidity density distribution
\ie\ the mean number of charged particles per single pp-collision and unit of pseudorapidity.
Only the per track corrections for background and tracking efficiency are needed.
For this distribution, at least one  VELO track is
required in the full forward $\eta$ range. Each of these elements of the analysis procedure
are discussed in subsequent subsections.

Hard interaction events are defined by requiring at least one long track with
$\pt >1$\gevc in the range $2.5 < \eta < 4.5$ where the detector has high
efficiency. The geometric acceptance is no longer independent of momentum and therefore the
distributions require an additional correction.

In this analysis
primary charged particles are defined as all particles for which the sum
of the ancestors' mean lifetimes is smaller than $10$~ps;
according to this definition the decay products of beauty and charm are
primary particles.

\subsection{Efficiency correction}
\label{subsec:efficiency}
The LHCb simulation is used to estimate the overall tracking and
selection efficiency as a function of pseudorapidity and azimuthal
angle $\phi$. As the \velo is outside the magnetic field region tracks
are straight lines and no study of acceptance as a function of
momentum is necessary. It is found that the efficiency (including
acceptance)  in the forward
region is typically greater than 90\% while it is at least 85\% in the
backward region.
Tracking efficiency depends weakly on the event track
multiplicity; this is taken into account in the evaluation of the
systematic error.

\subsection{Background contributions}
\label{subsec:background}
There are two main sources of background that can affect the
measurement of the multiplicity of charged particles:  secondary
particles misidentified as primary and fake
tracks. Other sources of background, such as beam-gas
interactions, are estimated to be negligible.

The correlation between the number of \velo hit clusters in an event and
its track multiplicity is in good agreement between the data and simulation,
indicating that the fraction of fake tracks is well understood. It is also
found that for each $\eta$ bin the multiplicity of fake tracks is
linearly dependent on the number of \velo clusters in the event.
Therefore it is possible to parameterise the fake contribution as a function of
\velo clusters using the Monte Carlo simulation.

The majority of secondary particles are produced in photon
conversions in the \velo material, and in the decay of long-lived
strange particles such as \KS and hyperons. While earlier LHCb measurements show that the production
of \KS is reasonably described by the Monte~Carlo generator~\cite{Aaij:2010nx},
there are indications that the production of \L particles is underestimated~\cite{Aaij:2011va}.
This difference is accounted for in the systematic error associated with the definition of
primary particles.

The fraction of secondary particles is estimated as a function of both $\eta$ and $\phi$.
In general, depending on the $\eta$  bin, the correction for non-primary particles
(from conversion and secondaries) changes the mean values of the
particle multiplicity distributions by $5-10\%$.

\subsection{Correction and unfolding procedure}
\label{subsec:corr_and_unfold}
The procedure consists of three steps; a background subtraction is made, followed by
an efficiency correction and finally a correction for pile-up. The procedure
is applied to all measured track multiplicity distributions in each of the different $\eta$ intervals.

In the first step, the distribution is corrected for fake tracks and non-primary particles.
A mean number of background tracks is estimated for each event based on the parameterizations described
in Section~\ref{subsec:background}. A PDF (probability density function) is built with this mean value
assuming a Poisson distribution for the number of background tracks. Hence, a PDF for the number of
prompt charged particles in a given event is then obtained. These per event PDFs are summed up and
normalized to obtain the
reconstructed prompt charged track multiplicity distribution \ie\ the fraction of events with ${\rm n_{tr}}$
tracks, ${\rm Prob(n_{tr})}$.

In the second step, the correction for the tracking efficiency is applied. For each $\eta$ bin a mean
efficiency, $\epsilon$, is calculated based on the per track efficiency as function of $(\eta, \phi)$. As explained
below, this is used to unfold the background-subtracted track multiplicity distribution, ${\rm Prob(n_{tr})}$,  to
obtain the underlying charged particle multiplicity distribution, ${\rm Prob(\tilde{n}_{ch})}$, where ${\rm \tilde{n}_{ch}}$
is the number of primary produced particles of all proton-proton collisions in an event.

For a given value of ${\rm \tilde{n}_{ch}}$, the probability
to observe ${\rm n_{tr}}$ reconstructed tracks given a reconstruction efficiency $\epsilon$
is described by the binomial distribution
\begin{equation}
p({\rm n_{tr}},{\rm \tilde{n}_{ch}},\epsilon) = \left(\begin{array}{c}
                      {\rm \tilde{n}_{ch}} \\
                      {\rm n_{tr}}
                    \end{array}\right)
                    (1-\epsilon)^{{\rm \tilde{n}_{ch}}-{\rm n_{tr}}}\epsilon^{{\rm n_{tr}}} .
\end{equation}

\noindent Hence, the observed track multiplicity distribution is given by
\begin{equation}
{\rm Prob(n_{tr})} = \sum_{ {\rm \tilde{n}_{ch}}=0}^{\infty} {\rm Prob(\tilde{n}_{ch})} \times p({\rm n_{tr}},{\rm \tilde{n}_{ch}},\epsilon) .
\end{equation}
The values for $ {\rm Prob(\tilde{n}_{ch})}$ are obtained by performing a fit to ${\rm Prob(n_{tr})}$.
The procedure has been verified using simulated data.

In the last step, the distributions are corrected for pile-up to obtain the charged particle multiplicity distributions of
single interaction events, ${\rm Prob(n_{ch})}$. This is done using an iterative procedure.
For low luminosity, $ {\rm Prob(\tilde{n}_{ch})}$  has mainly two contributions:
single proton-proton interactions and a convolution of two single proton-proton interactions.
The starting assumption is that the observed distribution is the single proton-proton interaction.
From this, the convolution term is calculated, and by subtracting it from the
observed distribution, a first order estimate for the single proton-proton distribution is obtained.
This can then be used to calculate again the convolution term and obtain a second order
estimate for the single proton-proton distribution. The procedure usually converges after the second iteration.
The pile-up correction typically changes the mean value of the particle
multiplicity distributions by $3-4\%$. It was checked that the contribution
from pile-up events with more than two proton-proton collisions is negligible.

As mentioned before, no unfolding procedure is required for the charged particles pseudorapidity
density, only the per track corrections for background tracks and tracking efficiency are applied. The
distribution is then normalized to the total number of proton-proton collisions including pile-up collisions.
In the case of hard interactions, the pseudorapidity density distribution of
the pile-up collisions without the $\pt$ cut is first subtracted. Finally, the
distribution is normalized to the total number of hard collisions.

\section {Systematic uncertainty}
\label{sec:syserrors}

\subsection{Efficiency}
\label{sec:trackerrors}
Studies based on data and simulation show that the error on the tracking efficiency for particles reaching
the tracking stations T1-T3 is $<3\%$~\cite{LHCb-PROC-2011-057}.
The tracking efficiency reduces for low-momentum ($\pt<50$\mevc) particles
due to interactions with the detector material and the residual magnetic
field in the \velo region. Since no momentum measurement exists for the reconstructed \velo tracks,
the estimate of a mean efficiency relies on the prediction of the LHCb Monte~Carlo model for the
contribution of low-momentum particles to the
total number of particles. The simulation predicts that in the forward
region the fraction of particles below a transverse momentum of
50\mevc is 2.4\%. The
corresponding average single track efficiency in this $\eta$ range
is measured to be 94\%. In the two extreme cases in which no
particles with \pt below
50\mevc were reconstructed or no such particles were produced
the average track efficiency
would be reduced by 1.2\% or increased by 1.1\%
respectively. Assuming a 25\% uncertainty on the number of low
momentum particles, as suggested by
the comparison between the measured particle multiplicity
and Monte~Carlo prediction, the additional contribution to the track
efficiency uncertainty is $<$ 1\%.
Adding this to the $3\%$ track reconstruction uncertainty, gives an overall $4\%$ error on
the track efficiency used in the unfolding procedure. The systematic
error contribution is then estimated by unfolding the multiplicity
distributions varying the tracking efficiency by $\pm 4\%$.
\subsection{Non-primary particles}
\label{subsec:non_prompt}
The main systematic uncertainty on the contribution of non-primary particles arises
from the knowledge of the detector material (15\%). Two thirds of non-primary particles
are due to conversions of photons from $\pi^0$ decays, resulting in an
10\% uncertainty. The multiplicity of  $\pi^0$ scales with the charged multiplicity,
therefore no additional error is applied. Varying by $\pm 40\%$ the production of
$\Lambda$ results in an uncertainty of about 5\% on the non-primary contribution.
A pessimistic assumption of a 25\% underestimation of the non-prompt contribution would
change the mean and RMS values of the particle multiplicity distributions by $-2\%$,
which can be neglected compared to the tracking efficiency uncertainty of $4\%$.

\subsection{Pile-up}
\label{subsec:pile_up}
The pile-up corrections inherit a systematic uncertainty from the
determination of the mean number of visible interactions of $10\%$.
This correction to the pile-up fraction is small and is negligible
compared to the systematic uncertainty due to the track efficiency correction.

\section{Results}
\label{sec:results}
\begin{figure}[!htb]
 \centering
\begin{minipage}[b]{0.47\linewidth}
  \includegraphics[scale=0.47]{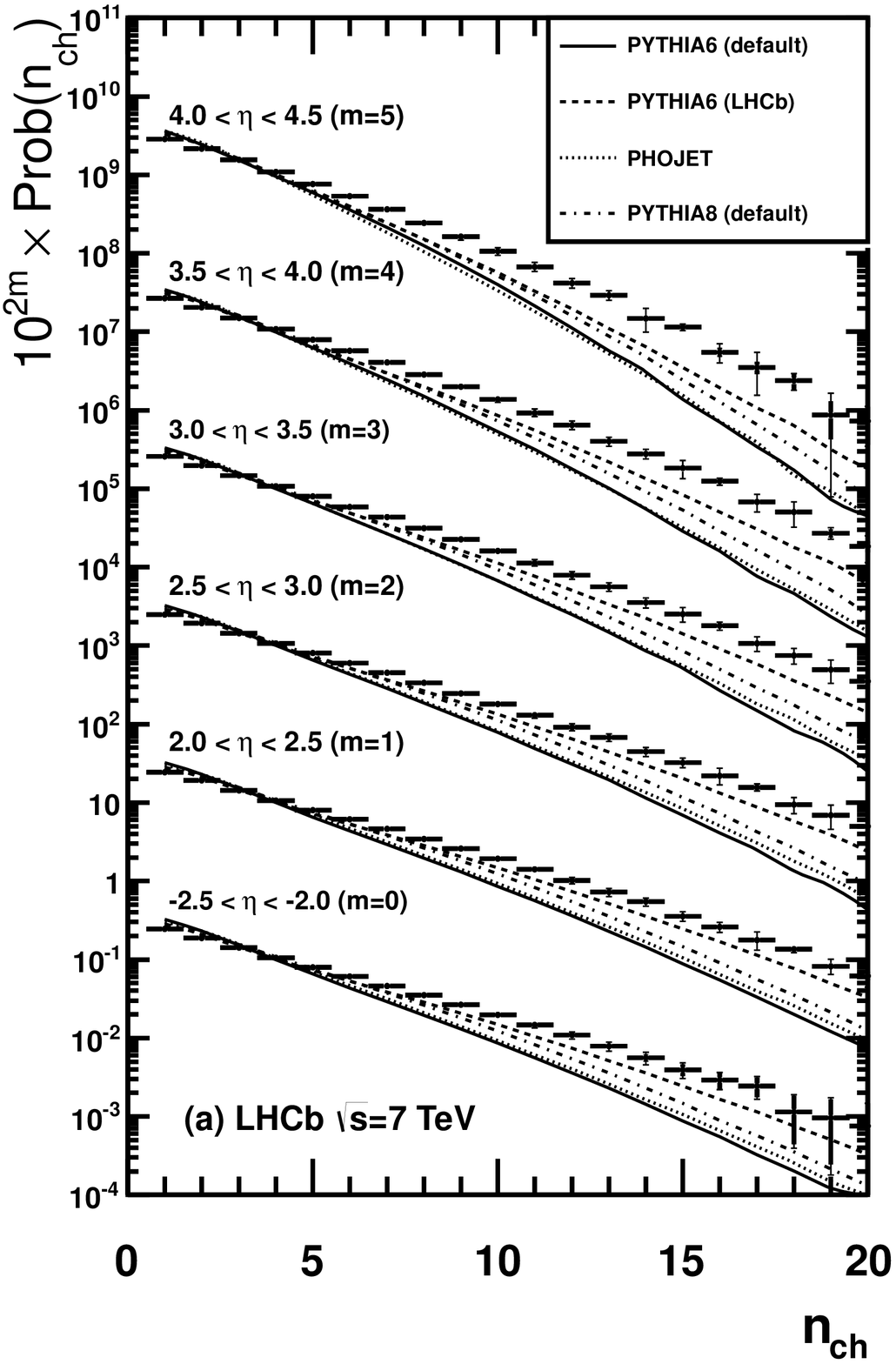}
\end{minipage}
\begin{minipage}[b]{0.47\linewidth}
  \includegraphics[scale=0.47]{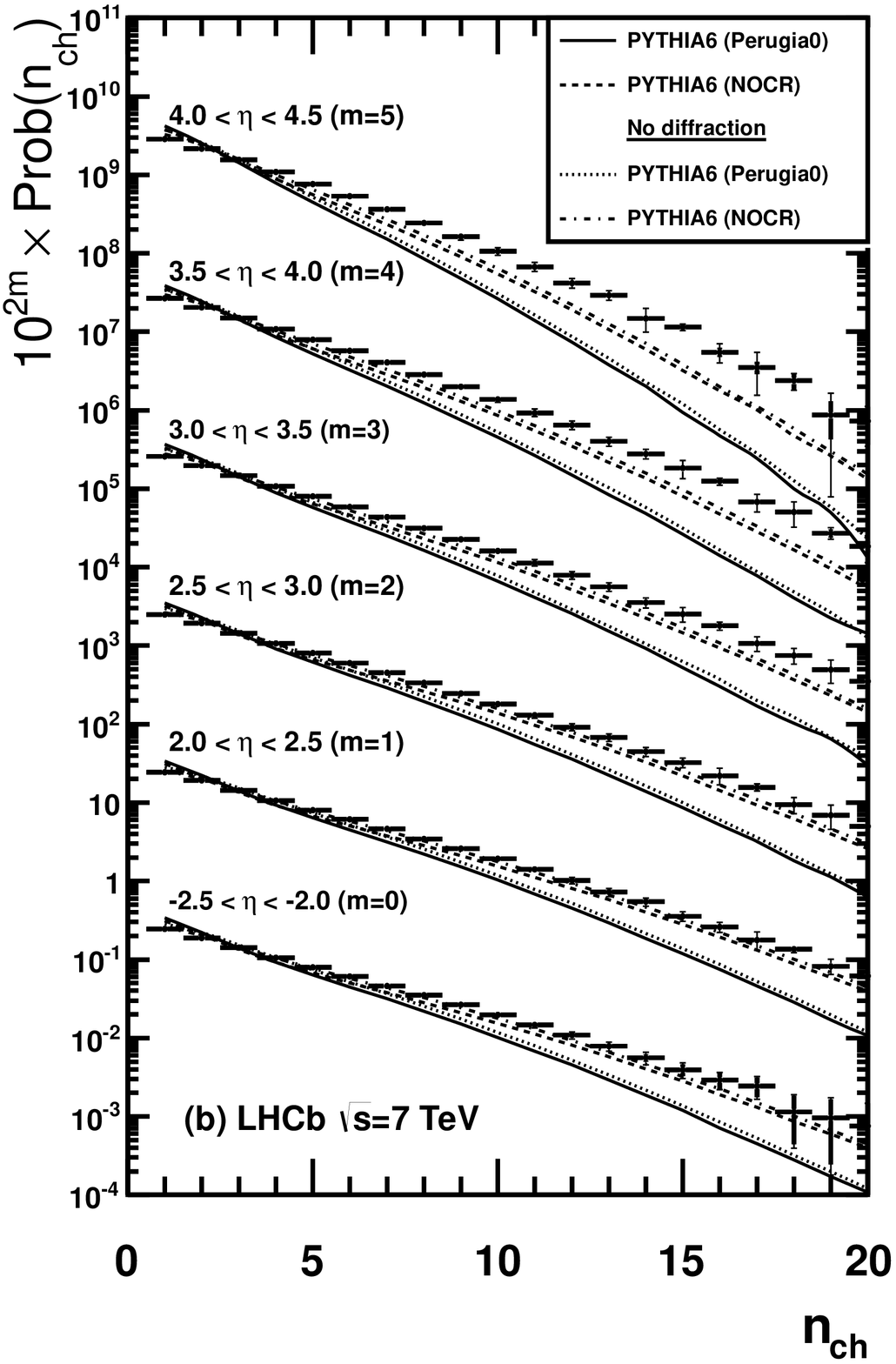}
\end{minipage}
  \caption{\small The multiplicity distribution in $\eta$ bins (shown as points with
    statistical error bars)
    with predictions of different  event generators.
    The inner error bar represents the statistical uncertainty and the outer error bar
represents the systematic and statistical uncertainty on the measurements. The data in both
figures are identical with predictions from \pythia6, \phojet and \pythia8 in (a) and
predictions of the \pythia6 Perugia tunes with and without diffraction in (b).}
   \label{fig:multiplicity}
\end{figure}

\begin{figure}[!htb]
 \centering
\begin{minipage}[b]{0.47\linewidth}
  \includegraphics[scale=0.47]{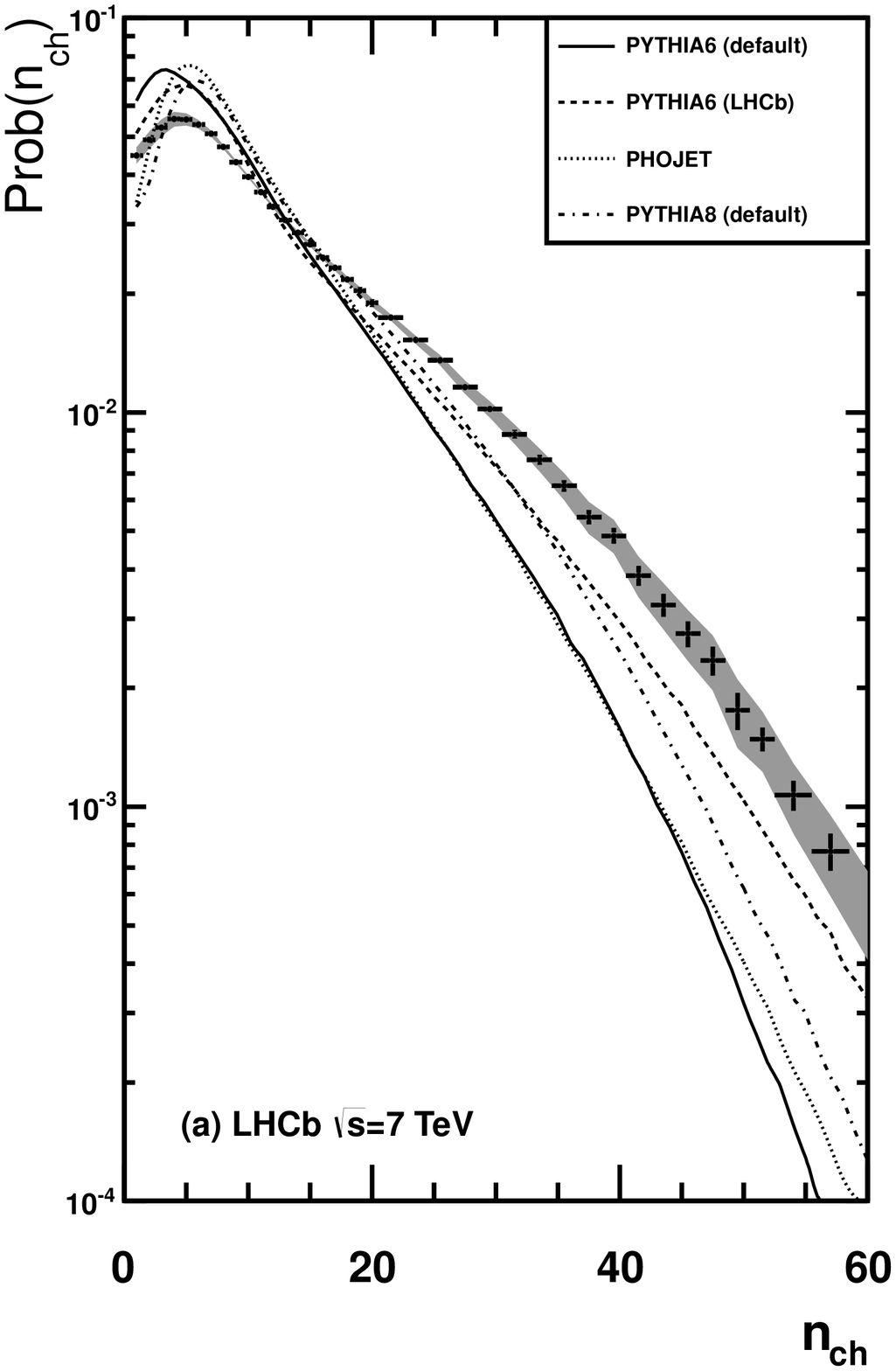}
\end{minipage}
\begin{minipage}[b]{0.47\linewidth}
  \includegraphics[scale=0.47]{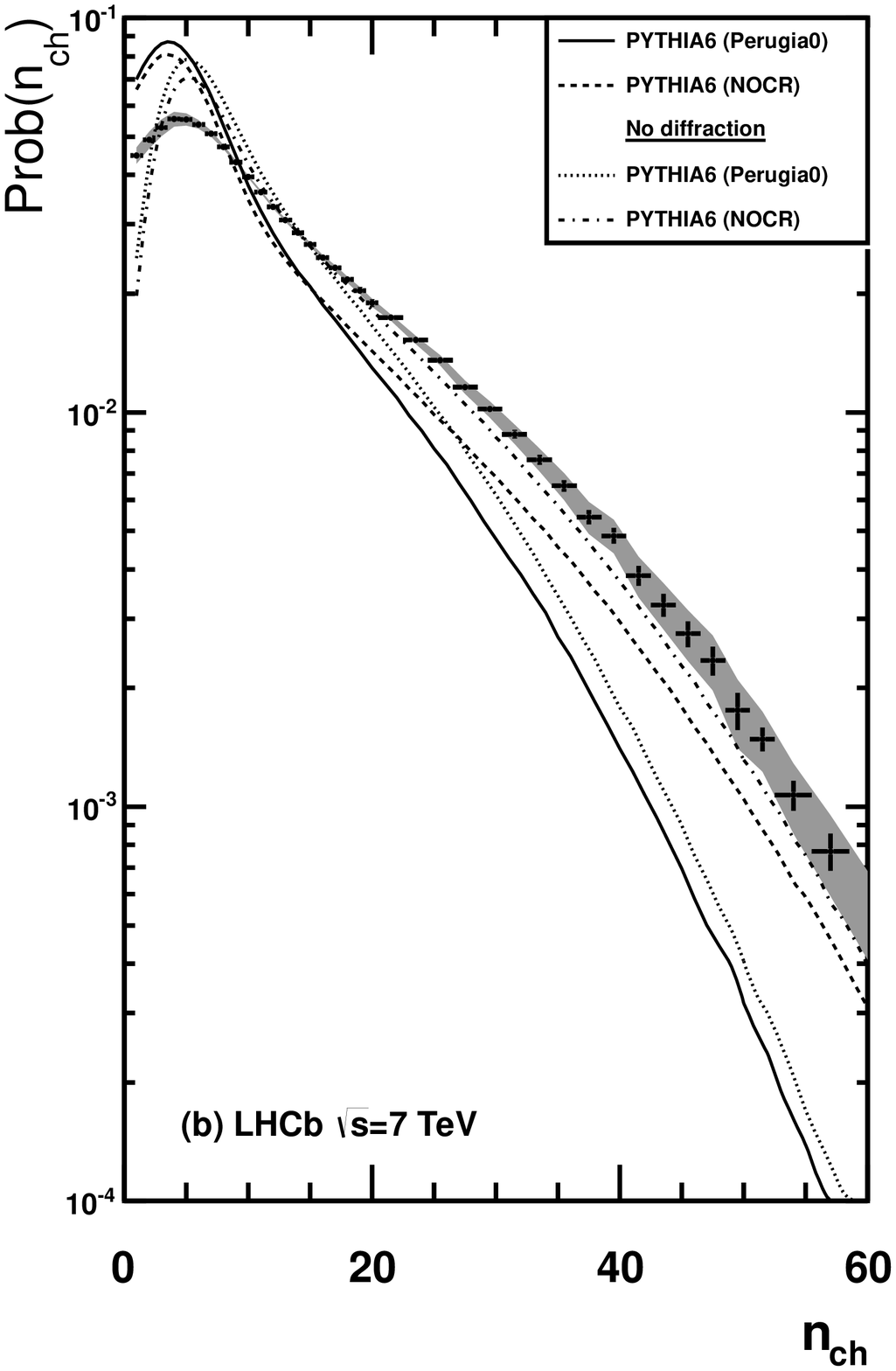}
\end{minipage}
  \caption{\small The multiplicity distribution in the forward $\eta$ range
   (shown as points with
    error bars)
    with predictions of different  event generators.
The shaded bands represent the total uncertainty on the measurements.
The data in both
figures are identical with predictions from \pythia6, \phojet and \pythia8 in (a) and
predictions of the \pythia6 Perugia tunes with and without diffraction in (b).}
   \label{fig:One_forPartMult}
\end{figure}

Figure~\ref{fig:multiplicity} shows the
unfolded charged particle multiplicity distribution
for different bins in
pseudorapidity, $\eta$.
Figure~\ref{fig:One_forPartMult} shows the multiplicity distributions
for the full forward range,
$2.0 < \eta < 4.5$.
There is a requirement of at least one track in the relevant $\eta$ range.
The distributions are compared to
several Monte~Carlo event generators.
\pythia 6.424 is
compared with the data for a number of tunes including the LHCb tuned
settings~\cite{LHCb-PROC-2010-056}. In particular the Perugia0 and PerugiaNOCR
tunings~\cite{Skands:2010ak} are shown.
In addition,
the \pythia 8.145 generator~\cite{Sjostrand:2007gs} was compared to
the data as well as \phojet 1-12.35~\cite{Phojet:1995a}.
In general all generators underestimate the multiplicity distributions,
with the LHCb tune giving the best description of the data; this tune
does not use data from the LHC.
The exclusion of the \pythia diffractive processes in the Perugia tunes, Figs.~\ref{fig:multiplicity}b
and~\ref{fig:One_forPartMult}b, also improves the description of the data, particularly in the
full forward region.

\begin{figure}[!htb]
  \centering
  \includegraphics [width=0.75\textwidth]{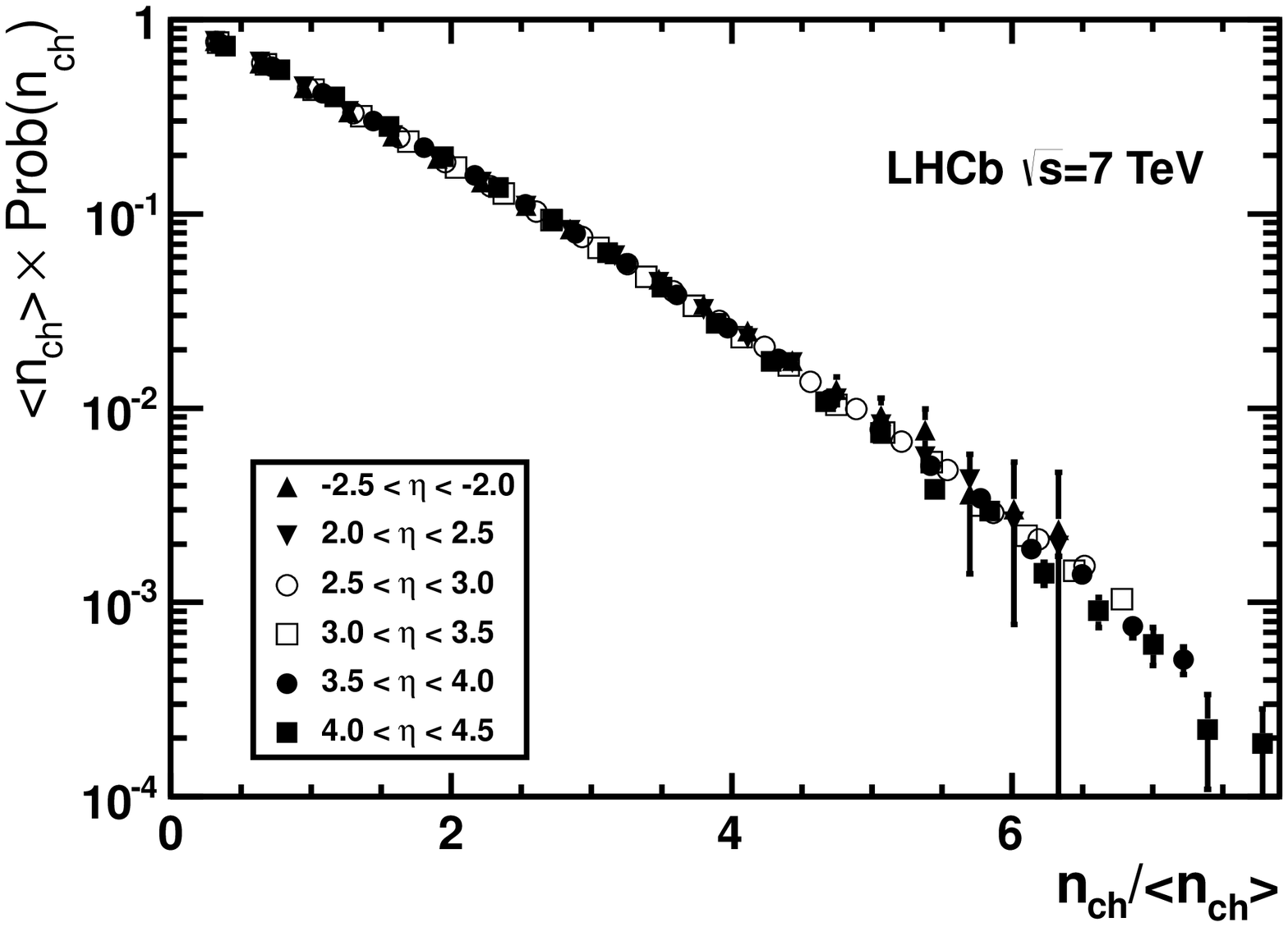}
  \caption{\small The KNO distributions in
    different bins of $\eta$. Only the
    the statistical uncertainties are shown.}
  \label{fig:kno}
\end{figure}

The Koba-Nielsen-Olesen (KNO) scaling variable~\cite{Koba:1972ng} has been used to compare the data
in the different  $\eta$ bins. Figure~\ref{fig:kno} shows the KNO scaled multiplicity
distributions, $\Psi(u)=\mean{{\rm n_{ch}}}\times {\rm Prob(n_{ch})}$ as a function of
$u=\frac{\rm n_{ch}}{\mean{\rm n_{ch}}}$.
As the multiplicity distributions measured are truncated the mean used
was extracted by fitting a negative binomial distribution.
It clearly shows that the distributions in the different $\eta$ bins are
equivalent. In particular this illustrates that when there is a requirement of at
least one track
in the $\eta$ bin the forward and backward regions $(2.0 < |\eta| < 2.5)$ are identical.

\begin{figure}[!htb]
  \centering
  \includegraphics[width=0.75\textwidth]{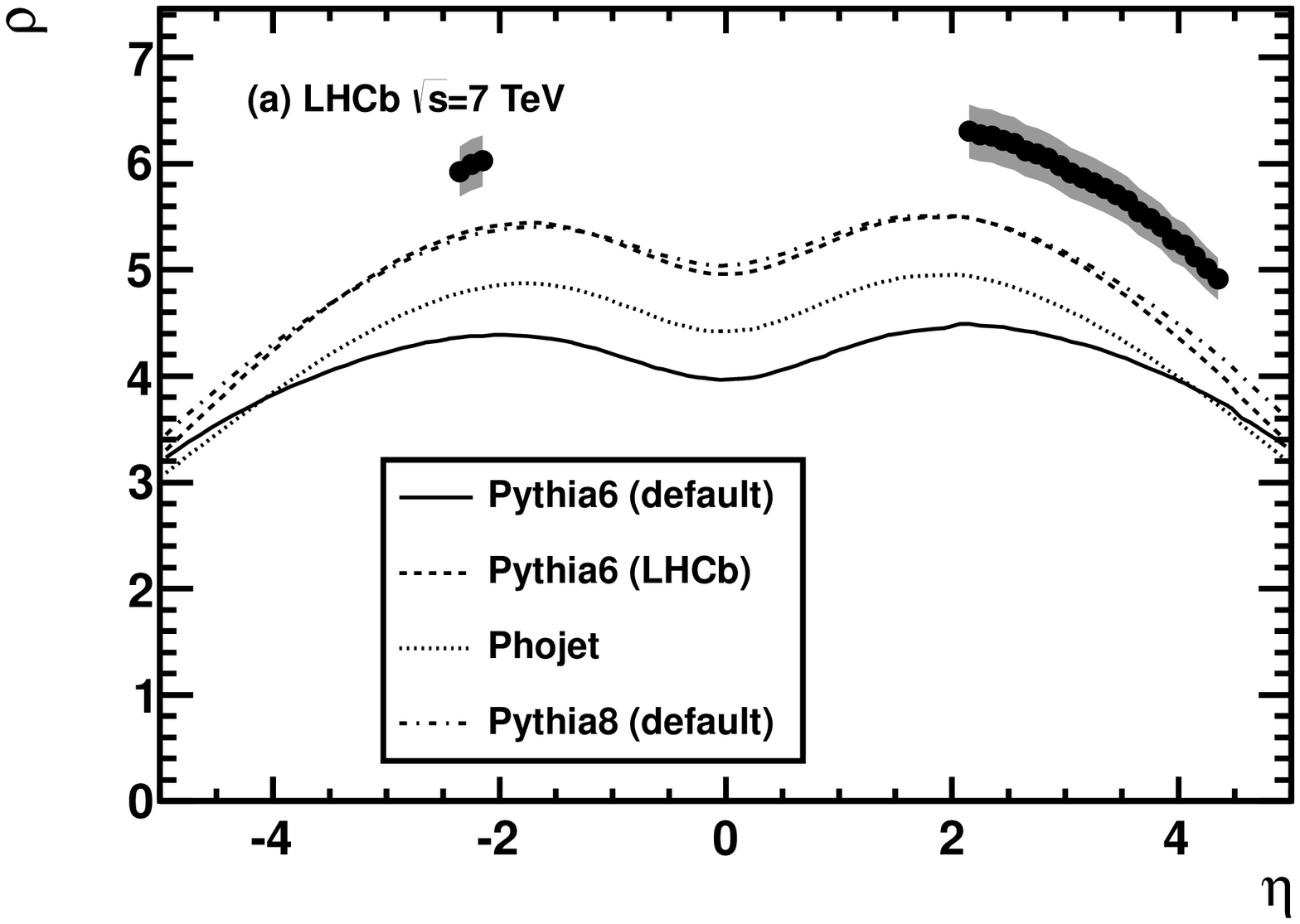}
  \includegraphics[width=0.75\textwidth]{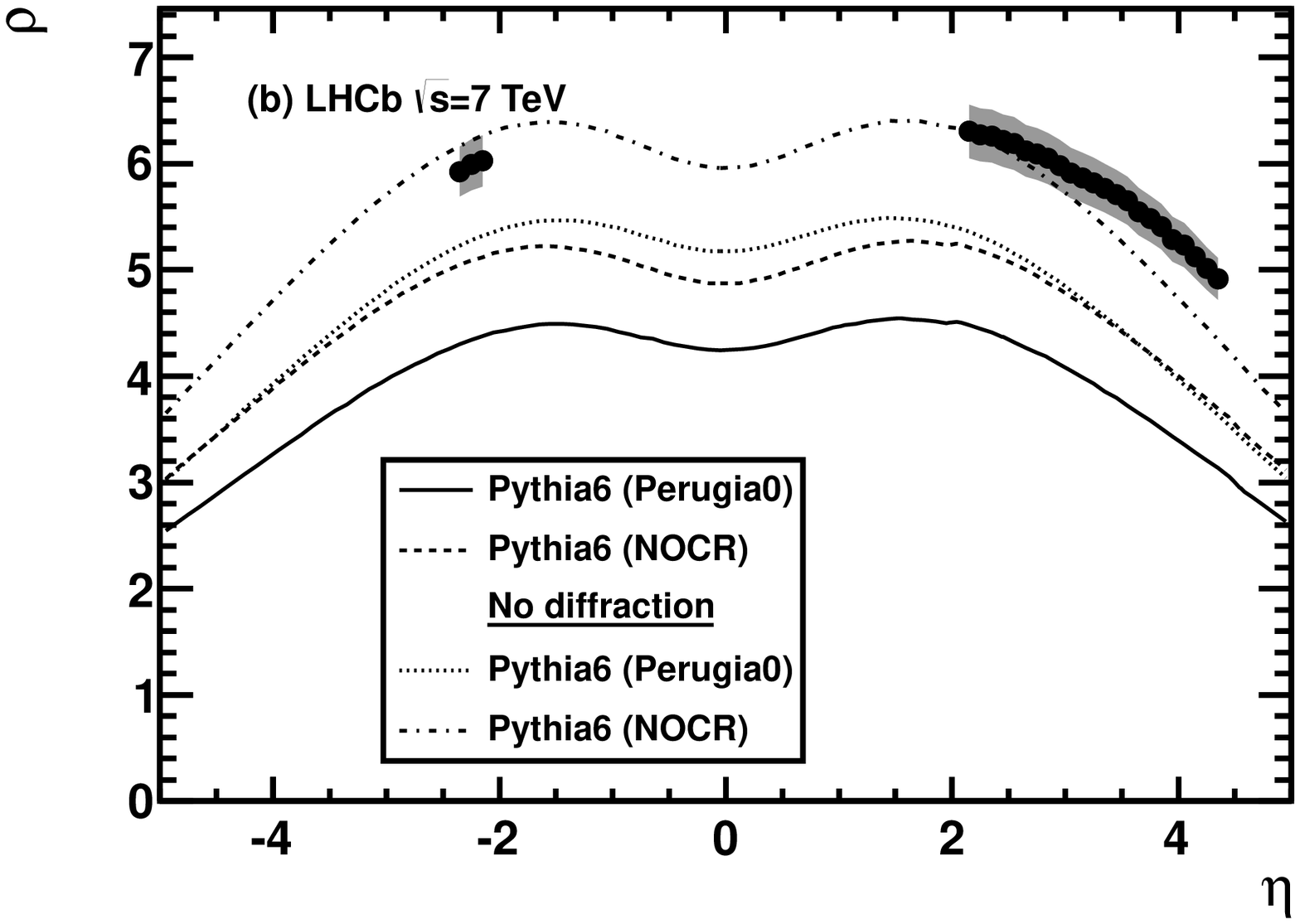}
  \caption{\small The charged particle densities as a function of
$\eta$ (shown as points with
    statistical error bars)
   and comparisons with predictions of event generators, as indicated in the key.
   The shaded bands represent the total uncertainty.
   The events are selected by requiring at least one charged particle in the range $2.0<\eta<4.5$.
The data in both
figures are identical with predictions from \pythia6, \phojet and \pythia8 in (a) and
predictions of the \pythia6 Perugia tunes with and without diffraction in (b).}
  \label{fig:mcrap}
\end{figure}

The charged particle pseudorapidity density, $\rho,$ is shown as a function of pseudorapidity in
Fig.~\ref{fig:mcrap}.
The data have a marked asymmetry between the
forward and backward region; this is a consequence of
the requirement of at least one track in the full forward $\eta$
range.
All models fail to describe the mean charged particle multiplicity per
unit of pseudorapidity.
The models, to varying degrees, also display the asymmetry but
in none of the models is this as large as in the data.
The effect on the predictions of excluding diffractive processes
is shown in Fig.~\ref{fig:mcrap}b using the Perugia tunes.
There is a better description of the $\eta$ distribution
in the backward directions but it still fails to describe the
forward-backward asymmetry.

\begin{figure}[!htb]
 \centering
\begin{minipage}[b]{0.47\linewidth}
  \includegraphics[scale=0.47]{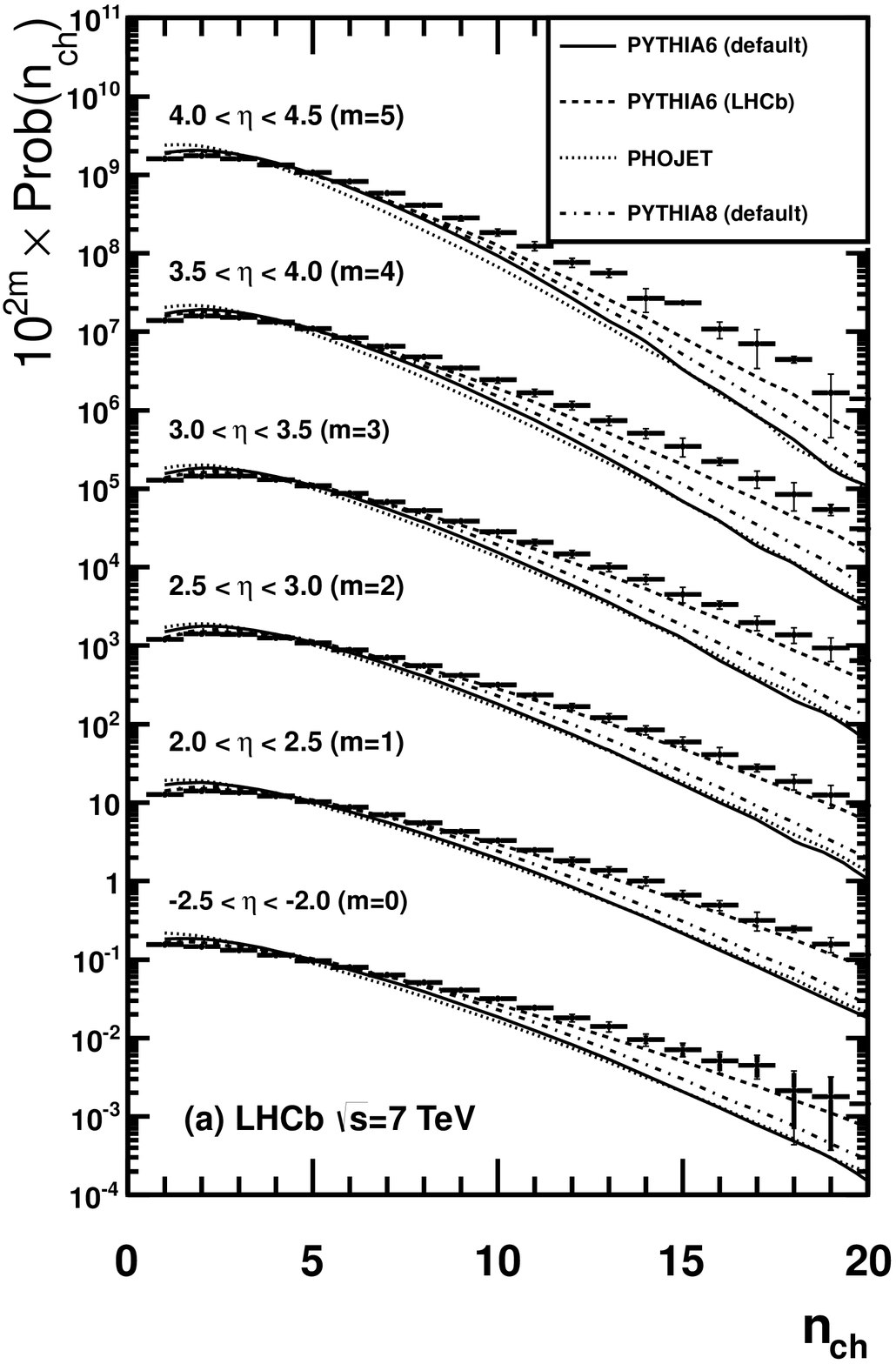}
\end{minipage}
\begin{minipage}[b]{0.47\linewidth}
   \includegraphics[scale=0.47]{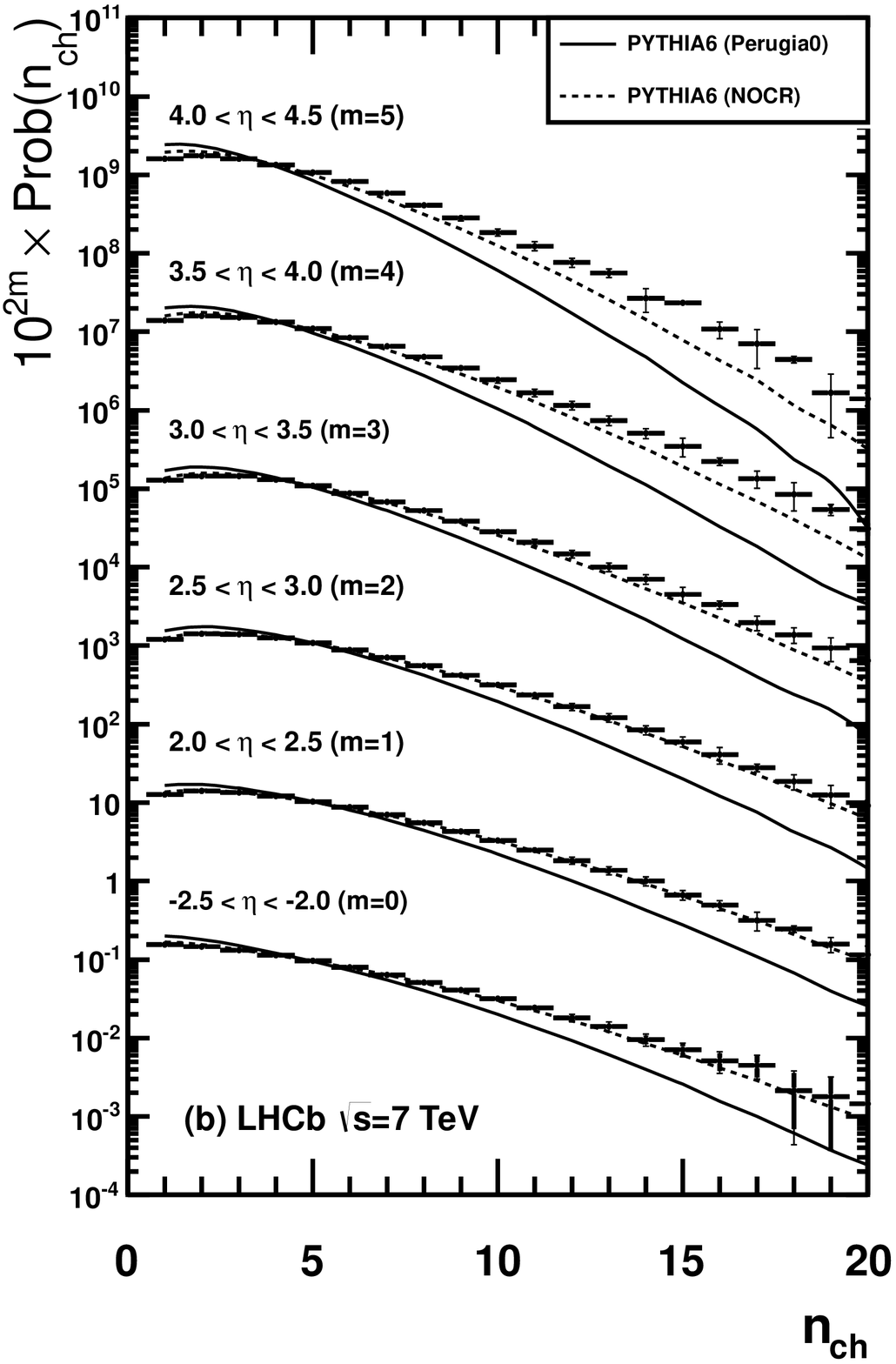}
\end{minipage}
  \caption{\small The multiplicity distribution in $\eta$ bins (shown as points with
    error bars)
    with predictions of different  event generators.
    The inner error bar represents the statistical uncertainty and the outer error bar
represents the systematic and statistical uncertainty on the measurements.
    The events have at least one track with a $\pt >1.0$\gevc in the
    pseudorapidity range $2.5 < \eta < 4.5$.
The data in both
figures are identical with predictions from \pythia6, \phojet and \pythia8 in (a) and
predictions of the \pythia6 Perugia tunes in (b).}
   \label{fig:multiplicity_hard}
\end{figure}

\begin{figure}[!htb]
 \centering
\begin{minipage}[b]{0.47\linewidth}
  \includegraphics[scale=0.47]{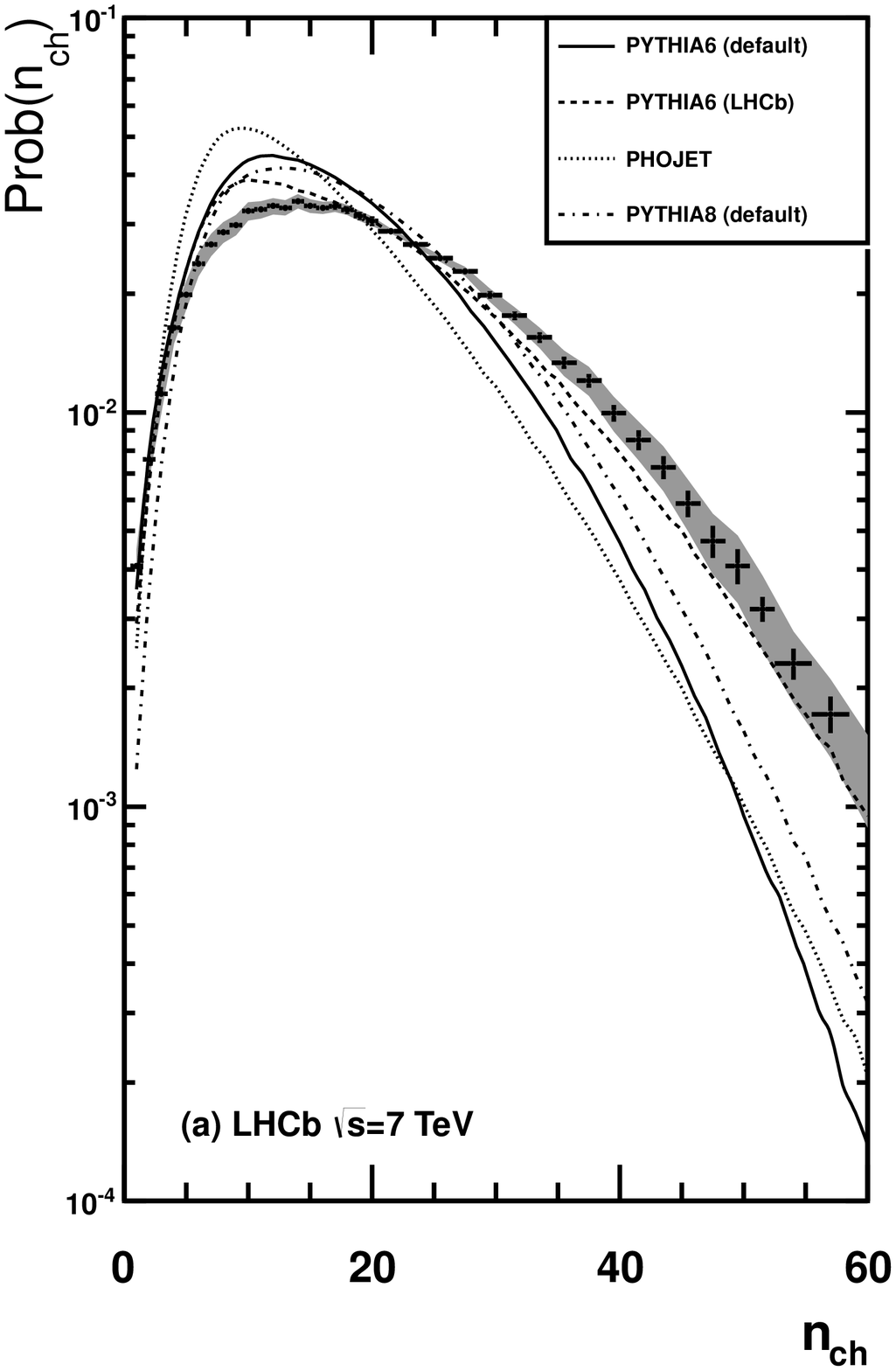}
\end{minipage}
\begin{minipage}[b]{0.47\linewidth}
   \includegraphics[scale=0.47]{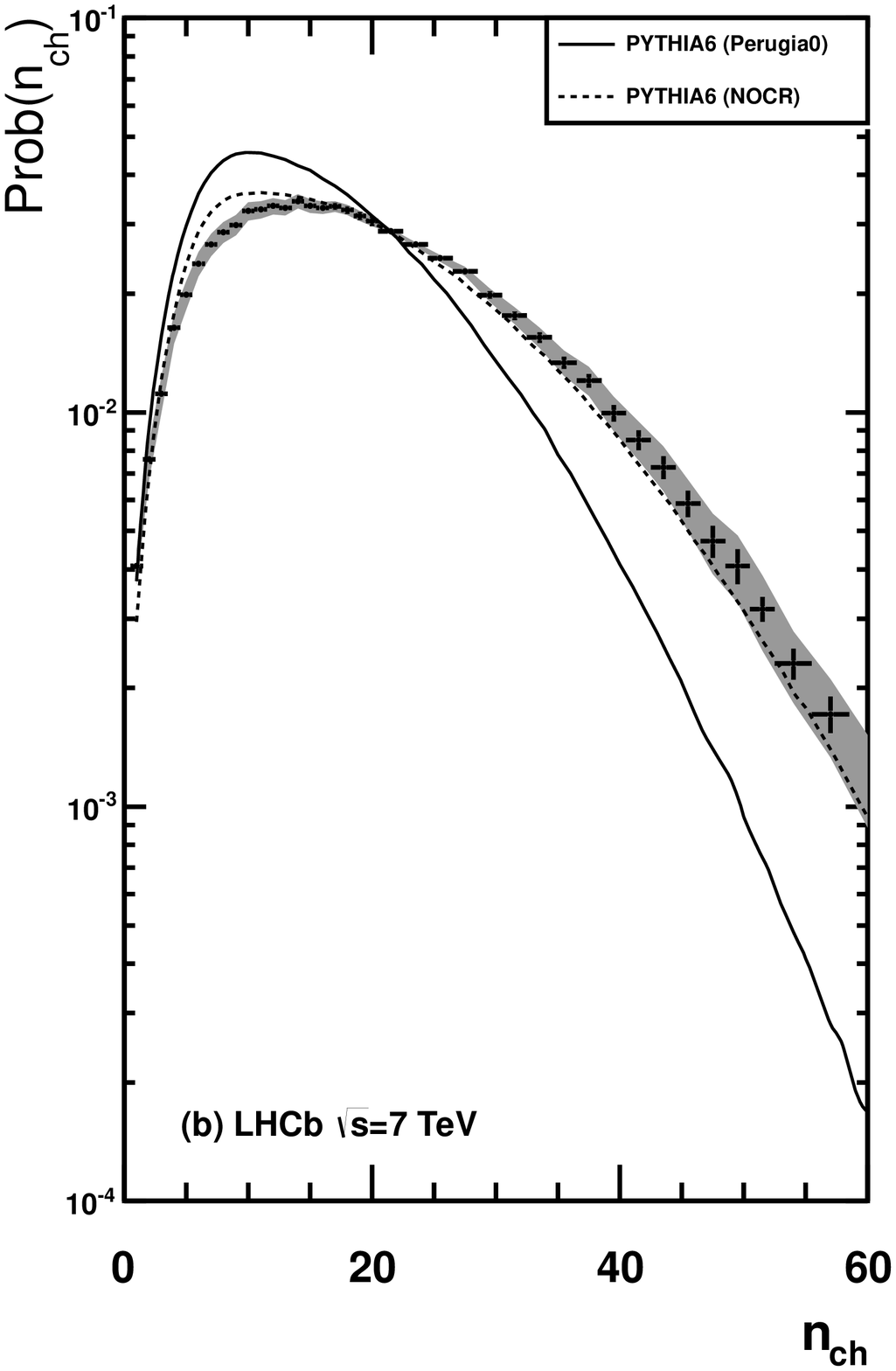}
\end{minipage}
  \caption{\small The multiplicity distribution in the forward
    $\eta$ range (shown as points with
    statistical error bars)
    with predictions of different  event generators.
    The shaded bands represent the total uncertainty.
    The events have at least one track with a $\pt >1.0$\gevc in the
    pseudorapidity range $2.5 < \eta < 4.5$.
The data in both
figures are identical with predictions from \pythia6, \phojet and \pythia8 in (a) and
predictions of the \pythia6 Perugia tunes in (b).}
   \label{fig:One_forPartMult_hard}
\end{figure}

A sample of hard QCD events were studied by ensuring at least one track in the
pseudorapidity range $2.5 < \eta < 4.5$ has a transverse momentum $\pt >1$\gevc.
In comparison to the data without this $\pt$ requirement,
the multiplicity distributions have larger high multiplicity tails,
see Figs.~\ref{fig:multiplicity_hard} and~\ref{fig:One_forPartMult_hard}.
The data are again compared to predictions of several event generators.
In general the predictions are in better
agreement than for the minimum bias data
but the pseudorapidity range $4.0 < \eta < 4.5$ remains
poorly described.  As the
\pt cut removes the majority of diffractive events from \pythia6 the
comparisons with and without diffraction are not shown.

\begin{figure}[!htb]
  \centering
  \includegraphics[width=0.75\textwidth]{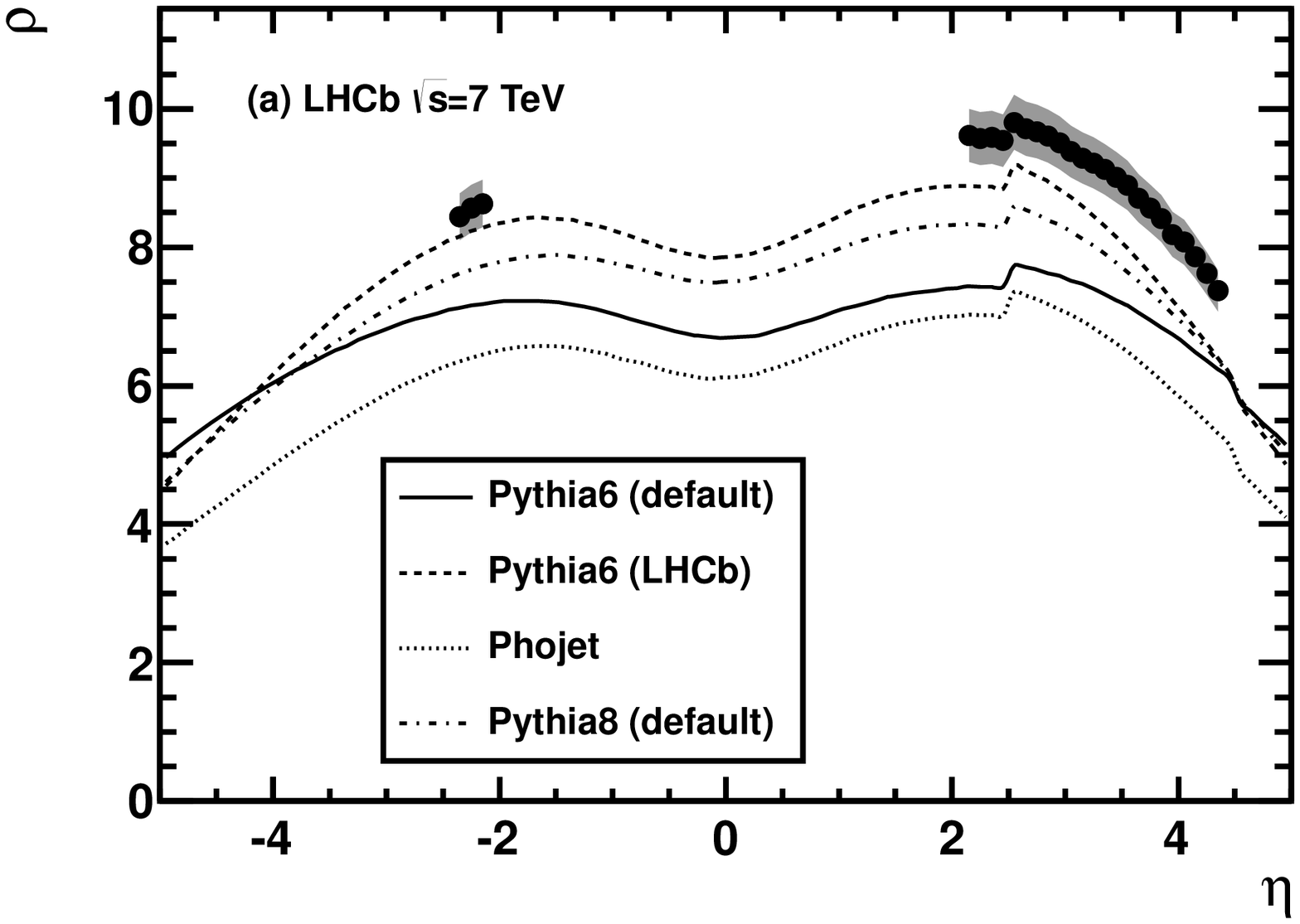}
  \includegraphics[width=0.75\textwidth]{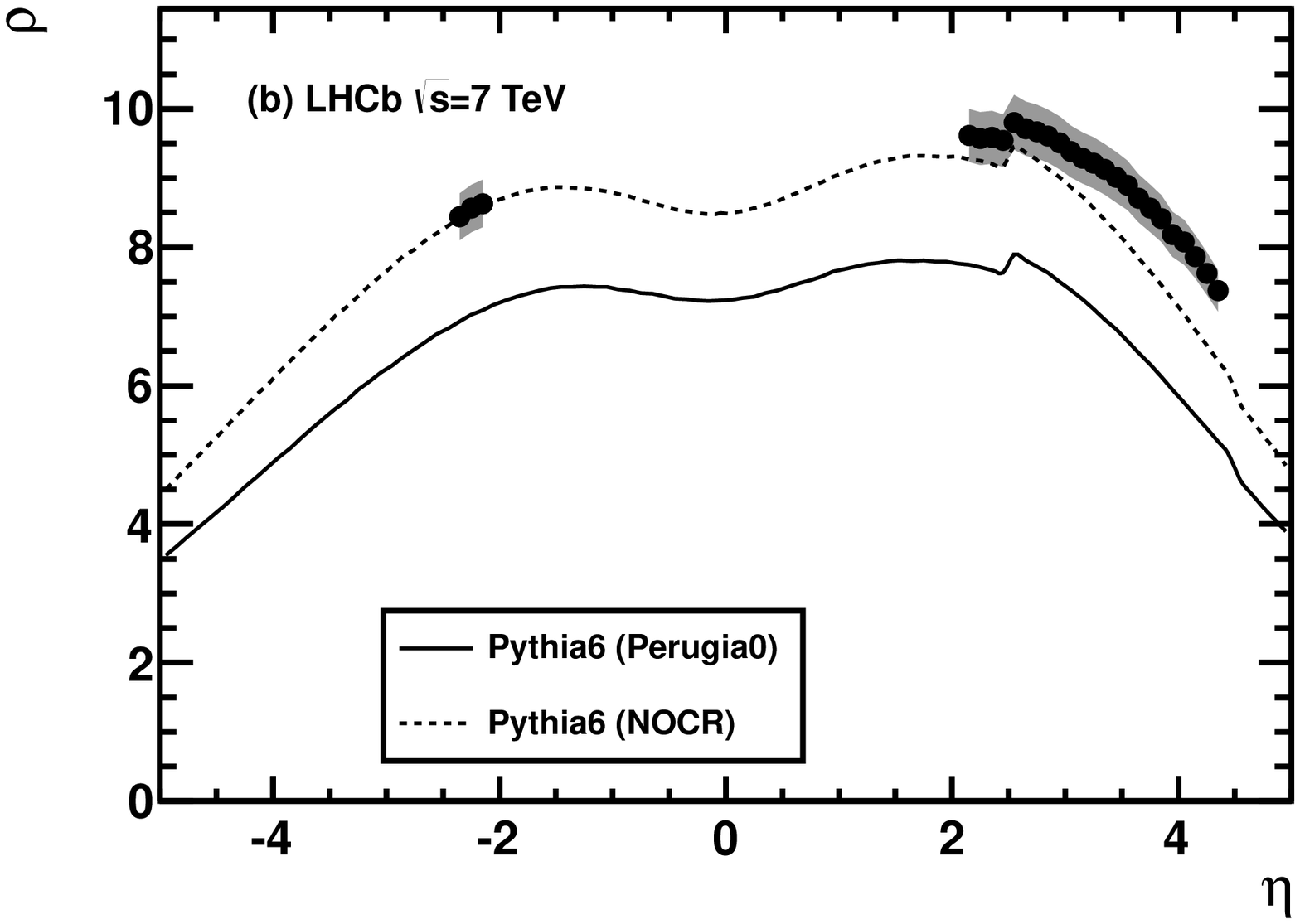}
  \caption{\small The data charged particle densities as a function of
$\eta$
(shown as points with
    statistical error bars)
and comparisons with predictions of event generators, as indicated in the key.
    The events have at least one track with a $\pt >1.0$\gevc in the
    pseudorapidity range $2.5 < \eta < 4.5$.
    The shaded bands represent the total uncertainty.}
  \label{fig:MCetapt}
\end{figure}

The charged particle density as a function of pseudorapidity for the
hard QCD sample is shown in
Fig.~\ref{fig:MCetapt}.
The discontinuity observed in the data at $\eta=2.5$ is an artefact of
the event selection for the hard events.
The asymmetry between the
forward and backward region is further amplified in this sample.
All models fail to describe the mean charged particle multiplicity per
unit of pseudorapidity.
The models, to varying degrees, also display the asymmetry but
never give an effect as large as the data.
The Perugia (NOCR) tune gives the best
description of the data in the backward direction but fails to
reproduce the size of the asymmetry.

\section{Summary}
\label{sec:summary}
The LHCb spectrometer acceptance, $2.0 < \eta <4.5,$ allows the forward
region to be probed at the LHC. The charged multiplicity distributions at
$\sqs = 7\tev$ are measured with and without a
\pt event selection, making use of the high efficiency of the LHCb \velo.
Several event generators are compared to the data; none are fully able to
describe the multiplicity distributions or the charged density
distribution as a function of $\eta$ in the LHCb acceptance.
 In general, the models
underestimate the charged particle production, in agreement with the
measurements in the central region at the LHC.

\clearpage

\section*{Acknowledgements}

\noindent We express our gratitude to our colleagues in the CERN accelerator
departments for the excellent performance of the LHC. We thank the
technical and administrative staff at CERN and at the LHCb institutes,
and acknowledge support from the National Agencies: CAPES, CNPq,
FAPERJ and FINEP (Brazil); CERN; NSFC (China); CNRS/IN2P3 (France);
BMBF, DFG, HGF and MPG (Germany); SFI (Ireland); INFN (Italy); FOM and
NWO (The Netherlands); SCSR (Poland); ANCS (Romania); MinES of Russia and
Rosatom (Russia); MICINN, XuntaGal and GENCAT (Spain); SNSF and SER
(Switzerland); NAS Ukraine (Ukraine); STFC (United Kingdom); NSF
(USA). We also acknowledge the support received from the ERC under FP7
and the Region Auvergne.

\clearpage
\raggedright
\bibliographystyle{LHCb}
\bibliography{EPJ}

\begin{appendices}
\newpage
\section{Tables of charged particle multiplicities}
\begin{table}[hbt]
\caption{\small \label{Table_multfit_back_bb} Charged particle multiplicity distribution in the pseudorapidity range $-2.5<\eta<-2.0$ for minimum bias events and for hard QCD events (see text). The first quoted uncertainty is statistical and the second is systematic.}
\begin{center} \begin{tabular}{|c|c|c|}
\hline
$ n_{\rm ch}$ & Prob.\ in min.\ bias  & Prob.\ in hard QCD \\
    &  events $\times 10^3$ &  events $\times 10^3$ \\ \hline
 $  1$  & $246.66\pm0.40\pm7.96$ &  $155.54\pm0.49\pm6.47$  \\
 $  2$  & $188.43\pm0.41\pm4.03$ &  $146.92\pm0.55\pm5.26$  \\
 $  3$  & $141.00\pm0.41\pm1.25$ &  $132.46\pm0.61\pm3.20$  \\
 $  4$  & $105.57\pm0.42\pm0.11$ &  $114.15\pm0.67\pm1.75$  \\
 $  5$  & $79.25\pm0.43\pm0.75$  &  $96.44\pm0.73\pm0.24$  \\
 $  6$  & $60.83\pm0.45\pm1.13$  &  $79.84\pm0.79\pm0.48$  \\
 $  7$  & $46.08\pm0.48\pm1.33$  &  $63.40\pm0.83\pm1.33$  \\
 $  8$  & $35.01\pm0.50\pm1.35$  &  $51.30\pm0.90\pm1.63$  \\
 $  9$  & $26.43\pm0.52\pm1.40$  &  $40.66\pm0.97\pm1.81$  \\
 $ 10$  & $19.75\pm0.55\pm1.36$  &  $31.50\pm1.02\pm1.86$  \\
 $ 11$  & $14.60\pm0.57\pm1.19$  &  $24.16\pm1.08\pm1.83$  \\
 $ 12$  & $10.82\pm0.59\pm1.00$  &  $18.03\pm1.12\pm1.64$  \\
 $ 13$  & $7.86\pm0.61\pm0.90$   &  $13.96\pm1.21\pm1.61$ \\
 $ 14$  & $5.57\pm0.63\pm0.86$   &  $9.56\pm1.19\pm1.28$  \\
 $ 15$  & $3.94\pm0.65\pm0.73$   &  $7.14\pm1.30\pm1.09$  \\
 $ 16$  & $2.90\pm0.67\pm0.37$   &  $5.10\pm1.29\pm1.11$  \\
 $ 17$  & $2.44\pm0.68\pm0.96$   &  $4.48\pm1.34\pm1.28$  \\
 $ 18$  & $1.14\pm0.70\pm0.61$   &  $2.13\pm1.43\pm2.03$  \\
 $ 19$  & $0.96\pm0.71\pm0.66$   &  $1.78\pm1.41\pm0.19$  \\
 $ 20$  & $0.75\pm0.72\pm0.27$   &  $1.46\pm1.44\pm0.60$  \\
\hline
\end{tabular}\end{center}
\end{table}

\begin{table}[!t]
\caption{\small \label{Table_multfit_for2025_bb} Charged particle multiplicity distribution in the pseudorapidity range $2.0<\eta<2.5$ for minimum bias events and for hard QCD events (see text). The first quoted uncertainty is statistical and the second is systematic. }
\begin{center} \begin{tabular}{|c|c|c|}
\hline
$ n_{\rm ch}$ & Prob.\ in min.\ bias  & Prob.\ in hard QCD \\
    &  events $\times 10^3$ &  events $\times 10^3$ \\ \hline
 $  1$  & $244.35\pm0.36\pm7.66$&  $126.88\pm0.38\pm6.57$  \\
 $  2$  & $191.00\pm0.33\pm4.02$&  $140.50\pm0.43\pm5.81$  \\
 $  3$  & $142.72\pm0.31\pm1.44$&  $133.83\pm0.44\pm3.91$  \\
 $  4$  & $106.75\pm0.28\pm0.10$&  $121.45\pm0.44\pm1.95$  \\
 $  5$  & $80.27\pm0.26\pm0.73$ &  $103.10\pm0.43\pm0.75$ \\
 $  6$  & $61.09\pm0.25\pm1.22$ &  $86.87\pm0.42\pm0.98$  \\
 $  7$  & $46.22\pm0.23\pm1.42$ &  $70.01\pm0.41\pm1.59$  \\
 $  8$  & $34.57\pm0.21\pm1.45$ &  $55.15\pm0.39\pm1.83$  \\
 $  9$  & $26.09\pm0.20\pm1.38$ &  $43.12\pm0.36\pm2.13$  \\
 $ 10$  & $19.30\pm0.18\pm1.34$ &  $32.71\pm0.34\pm2.20$  \\
 $ 11$  & $14.08\pm0.17\pm1.17$ &  $24.64\pm0.32\pm2.00$  \\
 $ 12$  & $10.17\pm0.16\pm1.07$ &  $18.25\pm0.29\pm1.80$  \\
 $ 13$  & $7.23\pm0.14\pm0.98$  &  $13.66\pm0.28\pm1.84$ \\
 $ 14$  & $5.43\pm0.13\pm0.82$  &  $9.97\pm0.25\pm1.52$  \\
 $ 15$  & $3.55\pm0.12\pm0.60$  &  $6.64\pm0.22\pm1.12$  \\
 $ 16$  & $2.60\pm0.11\pm0.40$  &  $4.91\pm0.21\pm0.78$  \\
 $ 17$  & $1.78\pm0.10\pm0.65$  &  $3.14\pm0.18\pm1.23$  \\
 $ 18$  & $1.35\pm0.09\pm0.28$  &  $2.45\pm0.17\pm0.47$  \\
 $ 19$  & $0.82\pm0.08\pm0.22$  &  $1.56\pm0.15\pm0.42$  \\
 $ 20$  & $0.62\pm0.07\pm0.19$  &  $1.15\pm0.13\pm0.34$  \\
\hline
\end{tabular}\end{center}
\end{table}

\begin{table}[!t]
\caption{\small \label{Table_multfit_for2530_bb} Charged particle multiplicity distribution in the pseudorapidity range  $2.5<\eta<3.0$ for minimum bias events and for hard QCD events (see text). The first quoted uncertainty is statistical and the second is systematic. }
\begin{center} \begin{tabular}{|c|c|c|}
\hline
$ n_{\rm ch}$ & Prob.\ in min.\ bias  & Prob.\ in hard QCD \\
    &  events $\times 10^3$ &  events $\times 10^3$ \\ \hline
 $  1$  & $249.37\pm0.35\pm7.88$ &  $121.02\pm0.36\pm6.72$  \\
 $  2$  & $194.45\pm0.33\pm4.11$ &  $140.71\pm0.41\pm6.20$  \\
 $  3$  & $144.53\pm0.29\pm1.39$ &  $138.90\pm0.42\pm4.26$  \\
 $  4$  & $107.18\pm0.27\pm0.10$ &  $125.71\pm0.41\pm2.10$  \\
 $  5$  & $80.42\pm0.24\pm0.89$  &  $108.13\pm0.40\pm0.34$ \\
 $  6$  & $60.29\pm0.22\pm1.34$  &  $87.75\pm0.37\pm1.24$  \\
 $  7$  & $45.03\pm0.20\pm1.53$  &  $70.69\pm0.35\pm1.85$  \\
 $  8$  & $33.53\pm0.18\pm1.55$  &  $55.79\pm0.33\pm2.31$  \\
 $  9$  & $24.75\pm0.16\pm1.46$  &  $42.12\pm0.30\pm2.40$  \\
 $ 10$  & $17.98\pm0.15\pm1.30$  &  $31.82\pm0.27\pm2.23$  \\
 $ 11$  & $12.98\pm0.13\pm1.23$  &  $23.37\pm0.25\pm2.10$  \\
 $ 12$  & $9.16\pm0.12\pm1.12$   &  $16.64\pm0.22\pm1.95$ \\
 $ 13$  & $6.74\pm0.11\pm0.87$   &  $12.07\pm0.19\pm1.52$ \\
 $ 14$  & $4.46\pm0.09\pm0.71$   &  $8.43\pm0.17\pm1.27$  \\
 $ 15$  & $3.23\pm0.08\pm0.47$   &  $5.97\pm0.15\pm0.88$  \\
 $ 16$  & $2.20\pm0.07\pm0.71$   &  $4.07\pm0.13\pm1.31$  \\
 $ 17$  & $1.57\pm0.06\pm0.32$   &  $2.78\pm0.11\pm0.52$  \\
 $ 18$  & $0.94\pm0.05\pm0.32$   &  $1.86\pm0.10\pm0.51$  \\
 $ 19$  & $0.69\pm0.05\pm0.33$   &  $1.26\pm0.09\pm0.56$  \\
 $ 20$  & $0.50\pm0.04\pm0.13$   &  $0.92\pm0.08\pm0.20$  \\
\hline
\end{tabular}\end{center}
\end{table}

\begin{table}[!t]
\caption{\small \label{Table_multfit_for3035_bb} Charged particle multiplicity distribution in the pseudorapidity range $3.0<\eta<3.5$  for minimum bias events and for hard QCD events (see text). The first quoted uncertainty is statistical and the second is systematic.  }
\begin{center} \begin{tabular}{|c|c|c|}
\hline
$ n_{\rm ch}$ & Prob.\ in min.\ bias  & Prob.\ in hard QCD \\
    &  events $\times 10^3$ &  events $\times 10^3$ \\ \hline
 $  1$  & $257.54\pm0.36\pm8.38$ &  $128.89\pm0.38\pm7.33$  \\
 $  2$  & $199.12\pm0.33\pm4.08$ &  $145.79\pm0.41\pm6.39$  \\
 $  3$  & $147.50\pm0.30\pm1.23$ &  $145.41\pm0.43\pm4.13$  \\
 $  4$  & $108.21\pm0.27\pm0.31$ &  $130.01\pm0.42\pm2.16$  \\
 $  5$  & $79.83\pm0.24\pm1.10$  &  $109.73\pm0.41\pm0.44$ \\
 $  6$  & $58.83\pm0.22\pm1.50$  &  $87.48\pm0.38\pm1.58$  \\
 $  7$  & $43.25\pm0.20\pm1.67$  &  $67.91\pm0.35\pm2.16$  \\
 $  8$  & $31.48\pm0.18\pm1.64$  &  $52.94\pm0.32\pm2.50$  \\
 $  9$  & $22.72\pm0.16\pm1.48$  &  $38.50\pm0.29\pm2.43$  \\
 $ 10$  & $16.12\pm0.14\pm1.28$  &  $28.21\pm0.26\pm2.21$  \\
 $ 11$  & $11.37\pm0.13\pm1.19$  &  $20.63\pm0.24\pm2.17$  \\
 $ 12$  & $7.89\pm0.11\pm1.07$   &  $14.74\pm0.21\pm1.83$ \\
 $ 13$  & $5.63\pm0.10\pm0.81$   &  $10.02\pm0.18\pm1.45$ \\
 $ 14$  & $3.54\pm0.08\pm0.67$   &  $7.00\pm0.16\pm1.02$  \\
 $ 15$  & $2.53\pm0.07\pm0.71$   &  $4.49\pm0.13\pm1.37$  \\
 $ 16$  & $1.79\pm0.06\pm0.38$   &  $3.33\pm0.12\pm0.64$  \\
 $ 17$  & $1.07\pm0.06\pm0.29$   &  $1.96\pm0.10\pm0.53$  \\
 $ 18$  & $0.75\pm0.05\pm0.17$   &  $1.38\pm0.09\pm0.32$  \\
 $ 19$  & $0.49\pm0.04\pm0.22$   &  $0.94\pm0.08\pm0.43$  \\
 $ 20$  & $0.35\pm0.04\pm0.10$   &  $0.65\pm0.07\pm0.17$  \\
\hline
\end{tabular}\end{center}
\end{table}

\begin{table}[!t]
\caption{\small \label{Table_multfit_for3540_bb} Charged particle multiplicity distribution in the pseudorapidity range $3.5<\eta<4.0$ for minimum bias events and for hard QCD events (see text). The first quoted uncertainty is statistical and the second is systematic.  }
\begin{center} \begin{tabular}{|c|c|c|}
\hline
$ n_{\rm ch}$ & Prob.\ in min.\ bias  & Prob.\ in hard QCD \\
    &  events $\times 10^3$ &  events $\times 10^3$ \\ \hline
 $  1$  & $268.35\pm0.37\pm8.77$  & $139.99\pm0.39\pm7.61$  \\
 $  2$  & $206.16\pm0.34\pm4.00$  & $158.42\pm0.44\pm6.72$  \\
 $  3$  & $150.62\pm0.31\pm0.98$  & $151.42\pm0.45\pm4.01$  \\
 $  4$  & $108.81\pm0.28\pm0.56$  & $133.07\pm0.44\pm1.67$  \\
 $  5$  & $78.99\pm0.25\pm1.35$   & $110.17\pm0.42\pm0.92$ \\
 $  6$  & $56.92\pm0.22\pm1.77$   & $84.74\pm0.38\pm1.91$  \\
 $  7$  & $40.49\pm0.20\pm1.81$   & $65.65\pm0.36\pm2.61$  \\
 $  8$  & $28.60\pm0.18\pm1.68$   & $48.06\pm0.32\pm2.71$  \\
 $  9$  & $19.98\pm0.16\pm1.46$   & $34.60\pm0.29\pm2.49$  \\
 $ 10$  & $13.79\pm0.14\pm1.30$   & $24.49\pm0.26\pm2.26$  \\
 $ 11$  & $9.31\pm0.12\pm1.18$    & $16.62\pm0.22\pm2.05$ \\
 $ 12$  & $6.48\pm0.11\pm0.94$    & $11.50\pm0.19\pm1.51$ \\
 $ 13$  & $4.02\pm0.09\pm0.68$    & $7.40\pm0.17\pm1.18$  \\
 $ 14$  & $2.80\pm0.08\pm0.41$    & $5.09\pm0.15\pm0.75$  \\
 $ 15$  & $1.82\pm0.07\pm0.64$    & $3.48\pm0.13\pm1.27$  \\
 $ 16$  & $1.24\pm0.06\pm0.28$    & $2.23\pm0.11\pm0.45$  \\
 $ 17$  & $0.68\pm0.05\pm0.25$    & $1.35\pm0.09\pm0.43$  \\
 $ 18$  & $0.50\pm0.04\pm0.21$    & $0.85\pm0.08\pm0.47$  \\
 $ 19$  & $0.27\pm0.04\pm0.05$    & $0.55\pm0.06\pm0.14$  \\
 $ 20$  & $0.18\pm0.03\pm0.08$    & $0.31\pm0.05\pm0.18$  \\
\hline
\end{tabular}\end{center}
\end{table}

\begin{table}[!t]
\caption{\small \label{Table_multfit_for4045_bb} Charged particle multiplicity distribution in the pseudorapidity range  $4.0<\eta<4.5$ for minimum bias events and for hard QCD events (see text). The first quoted uncertainty is statistical and the second is systematic.    }
\begin{center} \begin{tabular}{|c|c|c|}
\hline
$ n_{\rm ch}$ & Prob.\ in min.\ bias  & Prob.\ in hard QCD \\
    &  events $\times 10^3$ &  events $\times 10^3$ \\ \hline
 $  1$  & $284.08\pm0.40\pm9.11$  & $159.68\pm0.01\pm8.81$  \\
 $  2$  & $215.09\pm0.38\pm4.25$  & $174.85\pm0.01\pm6.65$  \\
 $  3$  & $155.18\pm0.35\pm0.72$  & $159.67\pm0.01\pm3.42$  \\
 $  4$  & $109.77\pm0.32\pm1.07$  & $135.15\pm0.01\pm0.61$  \\
 $  5$  & $76.74\pm0.29\pm1.76$   & $107.91\pm0.01\pm1.45$ \\
 $  6$  & $53.34\pm0.27\pm1.97$   & $82.45\pm0.01\pm2.49$  \\
 $  7$  & $36.49\pm0.24\pm1.93$   & $58.82\pm0.01\pm2.84$  \\
 $  8$  & $24.57\pm0.22\pm1.75$   & $41.25\pm0.01\pm2.75$  \\
 $  9$  & $16.30\pm0.20\pm1.50$   & $28.48\pm0.01\pm2.55$  \\
 $ 10$  & $10.63\pm0.17\pm1.25$   & $18.52\pm0.01\pm2.11$  \\
 $ 11$  & $6.76\pm0.15\pm1.00$    & $12.41\pm0.01\pm1.83$ \\
 $ 12$  & $4.20\pm0.13\pm0.70$    & $7.64\pm0.01\pm1.25$  \\
 $ 13$  & $2.92\pm0.12\pm0.57$    & $5.63\pm0.01\pm1.12$  \\
 $ 14$  & $1.48\pm0.10\pm0.86$    & $2.66\pm0.01\pm1.54$  \\
 $ 15$  & $1.15\pm0.09\pm0.33$    & $2.35\pm0.01\pm0.67$  \\
 $ 16$  & $0.55\pm0.07\pm0.21$    & $1.08\pm0.01\pm0.40$  \\
 $ 17$  & $0.35\pm0.06\pm0.28$    & $0.71\pm0.01\pm0.54$  \\
 $ 18$  & $0.24\pm0.05\pm0.12$    & $0.45\pm0.01\pm0.21$  \\
 $ 19$  & $0.09\pm0.04\pm0.13$    & $0.17\pm0.01\pm0.24$  \\
 $ 20$  & $0.07\pm0.04\pm0.02$    & $0.14\pm0.01\pm0.05$  \\
\hline
\end{tabular}\end{center}
\end{table}

\begin{table}[!t]
\caption{\small \label{Table_multfit_for2045_bb}  Charged particle multiplicity distribution in the pseudorapidity range $2.0<\eta<4.5$ for minimum bias events and for hard QCD events (see text). The first quoted uncertainty is statistical and the second is systematic.   }
\begin{center} \small \begin{tabular}{|c|c|c|}
\hline
$ n_{\rm ch}$ & Prob.\ in min.\ bias  & Prob.\ in hard QCD \\
    &  events $\times 10^3$ &  events $\times 10^3$ \\ \hline
 $  1$  & $51.23\pm0.16\pm2.05$ &  $5.38\pm0.09\pm0.45$    \\
 $  2$  & $56.09\pm0.18\pm2.35$ &  $10.02\pm0.14\pm1.10$   \\
 $  3$  & $60.21\pm0.20\pm2.38$ &  $14.69\pm0.17\pm2.04$   \\
 $  4$  & $63.32\pm0.21\pm2.81$ &  $21.62\pm0.23\pm2.16$   \\
 $  5$  & $63.18\pm0.23\pm1.82$ &  $26.22\pm0.26\pm1.88$   \\
 $  6$  & $61.39\pm0.24\pm1.14$ &  $31.38\pm0.31\pm1.94$   \\
 $  7$  & $58.08\pm0.25\pm0.57$ &  $35.13\pm0.35\pm1.87$   \\
 $  8$  & $53.81\pm0.26\pm0.24$ &  $37.72\pm0.39\pm1.67$   \\
 $  9$  & $49.25\pm0.27\pm0.32$ &  $39.37\pm0.43\pm2.27$   \\
 $ 10$  & $45.18\pm0.28\pm0.26$ &  $42.69\pm0.49\pm2.31$   \\
 $ 11$  & $41.36\pm0.29\pm0.28$ &  $43.07\pm0.53\pm1.37$   \\
 $ 12$  & $37.94\pm0.31\pm0.35$ &  $43.97\pm0.58\pm1.39$   \\
 $ 13$  & $35.09\pm0.32\pm0.30$ &  $43.52\pm0.63\pm1.71$   \\
 $ 14$  & $32.55\pm0.34\pm0.33$ &  $45.25\pm0.70\pm2.01$   \\
 $ 15$  & $30.48\pm0.36\pm0.43$ &  $43.98\pm0.75\pm0.86$   \\
 $ 16$  & $28.20\pm0.38\pm0.48$ &  $43.48\pm0.81\pm0.90$   \\
 $ 17$  & $26.55\pm0.40\pm0.40$ &  $43.85\pm0.89\pm0.74$   \\
 $ 18$  & $24.83\pm0.43\pm0.39$ &  $42.96\pm0.96\pm0.34$   \\
 $ 19$  & $23.26\pm0.45\pm0.39$ &  $41.47\pm1.02\pm0.24$   \\
 $ 20$  & $21.64\pm0.48\pm0.59$ &  $40.21\pm1.09\pm0.29$   \\
 $ 21$  & $19.87\pm0.19\pm0.46$ &  $37.97\pm0.43\pm0.51$   \\
 $ 23$  & $17.44\pm0.20\pm0.52$ &  $35.08\pm0.46\pm0.67$   \\
 $ 25$  & $15.49\pm0.21\pm0.76$ &  $32.39\pm0.51\pm0.87$   \\
 $ 27$  & $13.24\pm0.22\pm0.68$ &  $30.02\pm0.56\pm1.42$   \\
 $ 29$  & $11.63\pm0.23\pm0.60$ &  $26.14\pm0.57\pm1.54$   \\
 $ 31$  & $10.05\pm0.24\pm0.62$ &  $23.18\pm0.60\pm1.38$   \\
 $ 33$  & $8.66\pm0.25\pm0.62$  &  $20.40\pm0.63\pm1.45$  \\
 $ 35$  & $7.43\pm0.26\pm0.60$  &  $17.59\pm0.63\pm1.52$  \\
 $ 37$  & $6.19\pm0.26\pm0.72$  &  $15.85\pm0.66\pm1.88$  \\
 $ 39$  & $5.56\pm0.26\pm0.71$  &  $13.11\pm0.64\pm1.45$  \\
 $ 41$  & $4.40\pm0.25\pm0.62$  &  $11.22\pm0.64\pm1.32$  \\
 $ 43$  & $3.71\pm0.25\pm0.56$  &  $9.55\pm0.63\pm1.24$   \\
 $ 45$  & $3.14\pm0.24\pm0.44$  &  $7.74\pm0.59\pm1.27$   \\
 $ 47$  & $2.68\pm0.23\pm0.46$  &  $6.21\pm0.58\pm1.40$   \\
 $ 49$  & $2.00\pm0.22\pm0.49$  &  $5.38\pm0.54\pm1.09$   \\
 $ 51$  & $1.70\pm0.12\pm0.32$  &  $4.18\pm0.30\pm1.09$   \\
 $ 54$  & $1.22\pm0.11\pm0.24$  &  $3.04\pm0.27\pm0.69$   \\
 $ 57$  & $0.88\pm0.09\pm0.20$  &  $2.26\pm0.24\pm0.49$   \\
 $ 60$  & $0.63\pm0.08\pm0.15$  &  $1.58\pm0.21\pm0.45$   \\
\hline
\end{tabular}\end{center}
\end{table}

\end{appendices}
\end{document}